%% file: 0_bare_jrnl_.tex
\documentclass[lettersize,journal]{IEEEtran}
\usepackage{amsmath,amsfonts,bm}
\usepackage{algorithm}
\usepackage{algpseudocode}
\usepackage{array}
\usepackage{textcomp}
\usepackage{stfloats}
\usepackage{url}
\usepackage{verbatim}
\usepackage{graphicx}
\usepackage{xfrac}

\usepackage[
 font=footnotesize 
 ]{subfig}

\usepackage{cite}
\makeatletter
\def\endthebibliography{%
	\def\@noitemerr{\@latex@warning{Empty `thebibliography' environment}}%
	\endlist
}
\makeatother

\DeclareMathOperator*{\argmax}{argmax}   

\begin{document}

\title{RIS-Aided Bistatic Radar for Rapid NLOS Sensing in the Teraharetz Band}

\author{Furkan Ilgac, Emrah \v Ci\v sija, Aya Mostafa Ahmed, Musa Furkan Keskin, Aydin Sezgin, and Henk Wymeersch 

\thanks{This work is funded by the German Research Foundation (“Deutsche Forschungsgemeinschaft”) (DFG) under Project–ID 287022738 TRR 196 for Project S03. The conference version of this work published in 2021 IEEE Statistical Signal Processing  Workshop (SSP) \cite{Emrah}.}}



\maketitle

\begin{abstract} Observing targets shadowed by an obstacle with radars is a challenging problem also known as around-the-corner radar. As urban areas get populated with autonomous vehicles, this topic is expected to become an important topic for next-generation sensing applications to help their decision-making. Furthermore, the high directionality of the terahertz band, anticipated for use in future wireless systems, not only enables but also necessitates the effective management of the signal's multipath components. In this paper, we investigate a non-line-of-sight (NLOS) sensing problem at terahertz frequencies. To be able to observe the targets shadowed by a blockage, we propose a method using reconfigurable intelligent surfaces (RIS). We employ a bistatic multiple-input multiple-output (MIMO) radar system and scan the obstructed area with RIS using hierarchical codebooks (HCB). Moreover, we propose an iterative maximum likelihood estimation (MLE) scheme to yield the optimum sensing accuracy, converging to Cramer-Rao lower bound (CRLB). We take band-specific effects such as  diffraction and beam squint into account and show that these effects are relevant factors affecting localization performance in RIS-employed radar setups. We investigate the overall system performance under various conditions and derive the position error bounds under the given conditions. The simulation results show that in case of a blocked line-of-sight between the transmitter and the targets, the receiver can still localize all the targets with very good accuracy using the RIS. The initial estimates obtained by the HCBs can provide centimeter-level accuracy, and when the optimal performance is needed, at the cost of a few extra transmissions, the proposed iterative MLE method improves the accuracy to sub-millimeter accuracy, yielding the position error bound.
\end{abstract}

\begin{IEEEkeywords}
Reconfigurable intelligent surface, position error bound, Cram\'er-Rao bound, localization, terahertz, diffraction, beam squint
\end{IEEEkeywords}

\section{Introduction}
\input{1_Introduction}

\section{System Model} \label{chp:system}
\input{2_system_model}
\section{Radar Detection and Localization with Hierarchical Codebooks} \label{chp:algo}
\input{3_algorithm} 
\section{Maximum Likelihood Based Iterative Refinement} \label{chp:mle}
\input{3.1_refienment}
\section{Position Error Bound} \label{chp:peb}
  \input{4_crb}
\section{Results} \label{chp:res}
    \input{5_results}
\section{Conclusions} \label{chp:concl}
    \input{6_conclusions}
\section*{Appendix A}
    \input{Appendix_A}

\section*{Appendix B}
    \input{Appendix_B}


\bibliographystyle{IEEEtran}
\bibliography{references}





\vfill

\end{document}

%% file: 1_Introduction.tex
\IEEEPARstart{T}{he} terahertz band (0.1-10THz) poses great potential for various next-generation wireless applications. For instance, it is expected that 6G communications would be using large bandwidths available at these frequencies \cite{6G}. In addition, increased spatial resolution and highly directive and compact antenna arrays that are envisioned to be used in these bands would enable precise localization and position estimation, which would be an essential part of future wireless systems \cite{LOCC}. Moreover, since many molecules exhibit characteristic signatures at these frequencies, terahertz waves have been used in spectroscopy and material characterization applications \cite{Thz}. These properties of the terahertz band could enable new wireless applications where precise localization and characterization of the environment are remotely performed by a compact, mobile radar operating in the terahertz band \cite{MARIE}. However, this band is characterized by large path loss and molecular absorption which further degrades the channel. To remedy these losses, reconfigurable intelligent surfaces (RIS) have been proposed as a potential tool to shape the wireless environment \cite{Akyildiz}. These surfaces are planar structures consisting of reconfigurable reflecting elements usually spaced sub-wavelengths apart, used for changing the reflection behavior of the incoming electromagnetic wave by applying configurable phase shifts at each reflecting element. This achieves great flexibility to designers and consequently is getting increasing attention in many wireless applications \cite{ris1, yasemin,kevin,ris2}.

Recent works have also introduced the RIS into sensing applications. Thanks to the vastly increased number of channels, the additional information obtained with RIS has been utilized with data-driven methods to infer information about the environment. Researchers used machine learning methods on RIS-aided setups to perform human posture recognition \cite{ZhuHan}, radar imaging \cite{ZhuHan2}, and semantic segmentation \cite{ZhuHan3}. In other works, this idea is extended for the NLOS setting, for patient activity monitoring in a real-life environment \cite{nature}.

There has been significant work utilizing RIS as an integrated tool in the radar signal processing chain as well. In a foundational work \cite{Buzzi}, the mathematical models for various RIS-equipped radar system configurations are established, and a method for improving the target SNR with the help of RIS is investigated. In other works, researchers have proposed various beamformer design strategies for enhancing target detection performance in different radar setups, such as using active RIS \cite{Buzzi2} and MIMO radars \cite{RISMIMOCH}. Target reflectivity and Doppler velocity estimation based on CRLB minimization are performed in \cite{DopplerCRLB}. Similarly, a sensing algorithm derived from joint transmit and reflective beamformer design for CRLB minimization has been proposed \cite{EldarSensing}. In \cite{RISWaveform} waveform design and a corresponding target detection algorithm in a multi-RIS-aided setup is investigated.  The radar range equation for a RIS-aided monostatic setup in an NLOS scenario is derived in \cite{Aubry}, and an investigation of trade-offs in the system parameters is made. Similarly, in \cite{HenkNLOSSensing}, a monostatic RIS-aided dual function radio and communication system is examined under NLOS conditions, and a two-step procedure for solving generalized likelihood ratio test is proposed. In \cite{Ilgac} an experimental study and modelling of a RIS-aided NLOS RFID detection system operating arond 5GHz is conducted, The quadruple reflection that is required to occur in a monostatic NLOS sensing setup comes with severe degradation in signal power, and despite the help from the RIS, the system still has over 80dB path loss beyond 4 meters. Considering even higher path losses in the THz band motivates the practicality of the bistatic setups in NLOS sensing problems. By reducing the reflection order to three, and having a direct LoS between  the target and the receiver, the bistatic setups offer more potential in NLOS sensing.

To conclude, there has been growing research aiming to develop the best-performing beamformers for a given angle bin with respect to different metrics [15-22]. In most works, the angular grid is assumed to be fine enough, and the effect of practically feasible angular resolution on position estimation performance is not investigated. Although these beamformers can be computed offline for each angle bin, scanning the environment with high-performance, narrow beams poses a problem as it would require many transmissions to cover an entire scene. This is particularly a challenge in the next-generation sensing systems where high accuracy expectations require fine angular resolution \cite{sarieddin}. Moreover, intrinsic challenges of the terahertz band, such as diffraction, molecular re-radiation, and beam squint due to the wideband, have not been modeled and addressed in the aforementioned context.

In this paper, we employ a hierarchical codebook-based search initially introduced in \cite{Emrah} to accelerate the scanning process in a bistatic NLOS radar setup operating in the terahertz band. The primary achievements and original aspects of our manuscript can be summarized as follows:

\begin{itemize}
    \item \textbf{Terahertz radar signal model:} In this paper, we incorporate various terahertz effects into our signal model and observe their effect on radar localization performance. The terahertz band is characterized by peculiar effects that are not present in other wireless channels. Molecular absorption and re-radiation of this energy are two important phenomena in the terahertz channels. These factors cause extra path loss and interference in the received signal, which compels the use of large antenna arrays for compensation. Moreover, in most NLOS scenarios, the diffraction effects are usually overlooked. However, these effects become prominent in RIS-aided setups, as the intended signal path often has double or more reflections, significantly reducing signal strength. In these situations, the diffraction phenomena become a significant source of interference. In this paper, we model this effect and propose a method to alleviate its impact. Finally, since the terahertz band is envisioned to cover large bandwidths, distortions due to \textit{beam squint} effect on antenna beams must be modeled, causing degradation in localization performance. The algorithm described in this paper is not only able to incorporate the angular distortions that happen on the different frequency bands, but also benefit from it. To the best of our knowledge, such an advanced channel model and its artifacts have not been investigated in a radar localization context yet.
    
    \item \textbf{Codebook-based fast-scanning and refinement:} We propose a hierarchical codebook (HCB) based search that performs an iteratively finer search while omitting the empty angle sectors. This method drastically accelerates the localization process, which is especially valuable in terahertz applications, where the angular resolution (i.e., the search space) is expected to be high. Later, after obtaining the initial estimate with the hierarchical codebooks, the estimates are further refined by an iterative algorithm which formulates the RIS-aided multi-target as a maximum likelihood estimation problem. It has been observed that the refined algorithm obtains the Cramer-Rao lower bound.
\end{itemize}
The rest of this paper is organized as follows: Section \ref{chp:system} introduces the system model where the terahertz channel model and diffraction effects are discussed. Section \ref{chp:algo} introduces the method for detecting and localizing the targets. In Section \ref{chp:peb}, the optimal performance bound for the system is derived. In Section \ref{chp:res}, we present our findings, and in Section \ref{chp:concl} we summarize our conclusions.

\subsubsection*{Notation}
The mathematical notation used in this paper is as follows: Boldface is used for vectors $\textbf{a}$ (lower case), and matrices $\textbf{A}$ (upper case). The all-ones column vector of size N is indicated as $\textbf{1}_N$. The transpose, hermitian, conjugate and pseudoinverse operators are denoted by the symbols $(.)^{T}$, $(.)^{H} (.)^*$, and $(.)^{+}$, respectively.

%% file: 2_system_model.tex
\begin{figure}[ht]
    \centering
    \includegraphics[scale= 0.75]{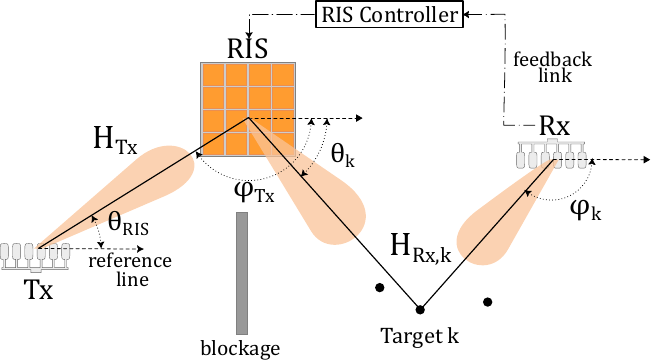}
    \caption{RIS-aided bistatic MIMO radar system with LOS blockage}
    \label{fig:system}
\end{figure}

An illustration of the system is given in Fig. \ref{fig:system}, which is essentially a bistatic MIMO radar setup with an obstacle between the transmitter and the receiver, preventing the line-of-sight (LoS). The strong attenuation and absorption characteristics of THz waves by most materials makes LoS path essential in THz wireless applications. In case LoS is not available due to any blocking objects, RIS has been proposed to maintain the surveillance of the shadowed area by constructing a controlled signal path. The transmitter (Tx), receiver (Rx), and RIS elements are arranged as uniform linear arrays (ULAs). The transmitter equipped with $N_{\text{Tx}}$ antennas transmits a signal to the RIS with $N_{\text{RIS}}$ discrete analog phase shifters. The RIS reflects the incoming signal behind the obstacle into the shadowed area. The target(s) in the mentioned area get(s) illuminated, and the receiver array with $N_{\text{Rx}}$ antennas collects the power reflected from the targets.

The reflections from the objects in the scene is modelled as stationary point targets. The  total number of targets (reflections) are denoted as $K$. The localization is performed on a two-dimensional (2D) global coordinate system. The positions of the transmitter, RIS, and the receiver are assumed to be known and denoted as $\textbf{p}_{\text{Tx}}$, $\textbf{p}_{\text{RIS}}$ and $\textbf{p}_{\text{Rx}}$ respectively. Without loss of generality, all of these arrays are assumed to be aligned along the same reference line. The probing signal steered as a sharp beam from the transmitter towards the RIS is specified as 
 \begin{equation} \label{eq:tx_signal}
 \textbf{x}(t,f) = 
 \sqrt{P_t} \frac{1}{N_{\text{Tx}}} 
 \textbf{a}_{{\text{Tx}}}(\theta_{\text{RIS}},f) 
 s(t,f),
 \end{equation} 
 where $P_t$ denotes the transmit power, $\textbf{a}_{\text{Tx}} \in  \mathbb{C}^{N_{\text{Tx}} \times 1}$ denotes the array steering vector directed towards the RIS with the  known angle $\theta_{\text{RIS}}$, and $s$ denotes the transmit signal. The time and frequency are given as $t$ and $f$, respectively. The elements of the transmit steering vector are in the form of $[\textbf{a}_{\text{Tx}}]_i = e^{-j2\pi \Delta \text{x} (i-1)\frac{f}{c}\cos{(\theta_\text{RIS}})}$, $\forall i = 1,\dots,N_{\text{Tx}}$ where $\Delta \text{x}$ is the antenna spacing and $c$ is the speed of light. Without loss of generality, the transmit signal is assumed to be  $s(t,f) = 1, \forall t$ in this work.
 
After the transmission, the signal propagates along the cascade of wireless channels, as illustrated in Fig. \ref{fig:system}. Due to the operating frequency, which is assumed to be $f_c \approx 300 $GHz for this work, some effects specific to the terahertz regime start to affect the signal propagation. These effects are discussed and modeled in the following section.
\subsection{Terahertz Propagation}
    \input{2.1_thz_model.tex}

\subsection{Diffraction Effects} \label{subs:DiffEff}
    \input{2.2_diff.tex} 
    
Putting intended signal paths and diffractive effects together, the received signal can be written as
\begin{equation} \label{eq:y}
\begin{split}
\Bar{\textbf{y}}[m] &= 
\sum_{k=1}^K 
( 
\sigma_k(m)
\textbf{H}_{\text{Rx,\textit{k}}}[m]
\bm{\Psi} 
\textbf{H}_{\text{Tx}}[m]
+\textbf{H}_{\text{diff,1}}[m]  \\
 &+
 \sigma_k(m) \textbf{H}_{\text{diff,2,k}}[m]
 )
\textbf{x}[m]
+\Bar{\textbf{w}}[m]\\       
&=
\sum_{k=1}^K
(\sigma_k(m)
l_{\text{tot}}(m)
\textbf{a}_{\text{Rx}}({\phi}_k,m) 
\textbf{a}_{\text{RIS}}({\theta}_k,m)^H
\bm{\Psi}\\
&\textbf{a}_{\text{RIS}}({\phi}_{\text{Tx}},m)
\textbf{a}_{\text{Tx}}({\theta}_{\text{RIS}},m)^H\\
&+l(\nu_1^m)
l_{\text{main}}^{\text{Tx-Rx}}(m,d)
\textbf{a}_{\text{Rx}}({\phi_{\text{diff}}^b,m}) 
\textbf{a}_{\text{Tx}}(\theta_{\text{diff}}^a,m)^H\\
&+\sigma_k(m)
l(\nu_2^m)
l_{\text{main}}^{\text{Tx-\textit{k}-Rx}}(m,d)
\textbf{a}_{\text{Rx}}({\phi}_k,m) 
\textbf{a}_{\text{Tx}}(\theta_{\text{diff}}^b,m) ^H
e^{j2\pi \beta_d}\\ 
&+
\textbf{H}_{\text{rad}}^{\text{diff}}
)\textbf{x}[m]
\Bar{\textbf{w}}[m]
\end{split}
\end{equation}
where $\sigma_k(m)$ denotes the radar cross section (RCS) of target $k$. The RCS varies significantly with frequency, necessitating the modeling of this relationship. For most objects, the RCS changes by about $8-10$ dBsm over a span of a few GHz \cite{rcs}. This variation also influences the choice of bandwidth for the localization algorithms. If the bandwidth of each sub-band is much smaller than the change in RCS, the RCS in each sub-band $\sigma_k(m)$ can be modelled as a constant value, which is Gaussian distributed across each frequency bin, with a mean of zero and a variance of 2.5. The phase shifts applied by the RIS is given by  $\bm{\Psi} = \textit{diag}(\psi) = \textit{diag}(e^{j\psi_1}), \dots e^{j\psi_{N_{\text{RIS}}}}$. The THz re-radiation-related components of the diffraction effects combined into single factor and denoted as
$\textbf{H}_{\text{rad}}^{\text{diff}} \in \mathbb{C}^{N_\text{Rx} \times N_\text{Rx}}$. Similarly, total path loss is also combined into a single factor as
$l_{\text{tot}}(m) = 
l_{\text{main}}^{\text{Tx-RIS}} (m)
l_{\text{main}}^{\text{RIS-\textit{k}-Rx}} (m)
$, and finally $\Bar{\textbf{w}}[m]$ denotes the additive white Gaussian noise with zero mean and variance $\sigma^2_w$, which is assumed to be known. 

Note that all array steering vectors in the expression above depends on frequency, hence written with frequency bin index $m$. On the other hand, the RIS does not perform baseband processing and, therefore, can only apply constant phase shift across all bins; this creates an artifact in the output of the RIS known as the \textit{beam squint}, which causes RIS beampattern to slightly divert on each frequency bin. This loss of focus causes a degradation in the SNR and localization performance, but methods incorporating the array factor-frequency relations, such as described in section \ref{chp:mle}, can alleviate its effect.

In addition, the effect of the first diffracted path $\textbf{H}_{\text{diff,1}}$ can be mitigated from the received signal by performing a background subtraction. To achieve this, a calibration signal needs to be generated by reflecting the transmitted signal off the RIS to a location confirmed to have no targets. This ensures that the received signal does not contain any target echoes, resulting in a signal primarily due to diffraction. Dropping the frequency band index $m$ for clarity, the calibration signal can be written as
\begin{equation} \label{eq:yd}
\textbf{y}_{\text{d}} = \sum_{k=1}^K 
\sigma_k \textbf{H}_{\text{Rx,\textit{k}}}  
\bm{\Psi}_{\text{Tx}} 
\textbf{H}_{\text{Tx}} 
\textbf{x}  +\textbf{H}_{\text{diff,2,k}}\textbf{x}
           +\textbf{H}_{\text{diff,1}}\textbf{x}
           +\textbf{w}_d,
\end{equation}
where $\bm{\Psi}_{\text{Tx}}$ represents the RIS phase profiles that focus the incoming beam back to the transmitter. This has been done to ensure no power leaks into blocked area, where targets exist. Also, $\textbf{w}_d$ denotes the noise term with same mean and variance as $\textbf{w}$.  The signal $\textbf{y}_d$ later can be subtracted from $\Bar{\textbf{y}}$ to suppress the effect of the first diffracted path. However, it must be noted that this method can not effectively cancel the second diffracted path, as it contains random phase terms. Therefore, the signal that is reaching the receiver over that path further contributes as interference to the system.

%% file: 2.1_thz_model.tex
First, the usual loss that is observed in wireless systems is the attenuation of the electromagnetic waves due to propagation in space. This phenomenon is called free space path loss (FPSL) and is characterized as 
\begin{equation}
 A_{\text{fspl}}(f,d) = \left( \frac{4\pi f d}{c}\right)^2,
\end{equation}
where  $f$ is the operating frequency, $c$ is the speed of the light and $d$ is the distance traveled by the signal \cite{proakis}.

Next, the molecular interactions become an important factor for wireless systems operating above 20GHz. If excited in their natural frequencies, the molecules in the air may absorb the incoming electromagnetic wave in the form of rotational-vibrational energy and cause additional loss, which is called atmospheric loss in wireless systems \cite{KOKKONIEMI}. Moreover, it has been demonstrated that the absorbed energy is re-radiated shortly after with the same frequency. This phenomenon is the source of self-induced noise and can cause significant interference in wireless systems operating on such high frequencies. To be able to model the channels more realistically, these interactions must be incorporated into the channel models. Air is a mixture of chemical compounds each of which has its own absorption spectra. Therefore, a combined absorption spectrum for the air can be written as a weighted sum of these compounds such as
\begin{equation}
 \kappa(f)= \sum_{i=1}^N \gamma_i \kappa_i(f),
\end{equation}
where $\gamma_i$ denotes the mole fraction of the molecule $i$, among the $N$ chemical compounds. $\kappa_i(f)$ denotes the absorption coefficient of the molecule $i$ at operating frequency $f$. Molecular absorption coefficients of many chemicals evaluated under different temperature and pressure conditions can be found in databases such as HITRAN and NIST Atomic Spectra \cite{reradiation}. After calculating the attenuation coefficient for the air mixture, the atmospheric loss can be modeled as
\begin{equation}
 A_{\text{atm}}(f,d) = e^{\kappa(f)d}. 
\end{equation}
Both of these effects attenuate the transmitted signal, hence the relation between the received power $P_{\text{r},\text{main}}(f,d)$ and the transmitted power $P_t(f)$ can be written as
\begin{equation}\label{eq:Pmain}
 P_{\text{r},\text{main}} (f,d) 
 = | h_{\text{main}}(f,d)|^2  P_t(f)
 = \frac{P_t(f)} {A_{\text{fspl}}(f,d)A_{\text{atm}}(f,d)}. 
\end{equation}
In above formulation, the main channel transfer function is denotes
\begin{equation} \label{eq:hmain}
 h_{\text{main}}(f,d) = l_{\text{main}}(f,d)   e^{j2\pi d f/c}.
\end{equation}
where 
\begin{equation} \label{eq:channel_gain}
l_{\text{main}}(f,d) =  \left( \frac{c}{4\pi f d}\right) e^{-k(f) d/2}     
\end{equation}
is the channel gain.
Note that the left-hand sides of the Equations (\ref{eq:Pmain}) and (\ref{eq:hmain})  are given with subscript "main" since they are modeling the main effects occurring in a terahertz channel. As mentioned before, molecular re-radiation is also an important source of interference in terahertz systems. Since the re-radiated signal is highly correlated, its effect can be modeled as interference. The relation between the re-radiated power $P_{\text{r},\text{int}} (f,d)$ and the transmit signal is given \cite{reradiation} 
 \begin{equation} \label{eq:reradiation}
    P_{\text{r},\text{rad}} (f,d) = (1 - e^{k(f)d})\left(\frac{c}{4\pi f d}\right)^2 P_t(f),
 \end{equation}
which yields the equivalent interference channel as
\begin{equation} \label{eq:hint}
 h_{\text{rad}}(f,d) =  
 \left(
 1- e^{-k(f)d} 
 \right)
 ^\frac{1}{2} 
 \left( 
 \frac{c}{4\pi f d}
 \right) 
 e^{j2\pi\beta_r}.
\end{equation}
Note that the phase of the re-radiated wave depends on the individual status of each molecule. To this end, a phase term has been introduced in the transfer function in the form of $e^{j2\pi\beta_r}$, where $\beta_r$ is assumed to be uniformly distributed in the range $ [0,1)$. Therefore, the overall channel model can be obtained by combining the channels given in (\ref{eq:hmain}) and (\ref{eq:hint}), yielding
\begin{equation} \label{eq:h}
    h(f,d) = h_{\text{main}}(f,d) + h_{\text{rad}}(f,d).
\end{equation}
The point-to-point single input single output (SISO) derived above can be extended into the MIMO case by introducing the array factors. Dividing the frequency spectrum into M separate coherent bins such as $f_m = f_c + \Delta f_m =  f_c + \frac{(2m-1-K)\text{BW}}{2M}$, where $\text{BW}$ denotes the total bandwidth, the channel between Tx and the RIS denoted by $\textbf{H}_{\text{Tx}} \in  \mathbb{C}^{N_{\text{RIS}} \times N_{\text{Tx}}}$ becomes
\begin{equation}
    \begin{split}
\textbf{H}_{\text{Tx}}[m] &=
l_{\text{main}}^{\text{Tx-RIS}}(m,d)
\textbf{a}_{\text{RIS}}(\phi_{\text{Tx}},m)
\textbf{a}_{\text{Tx}}(\theta_{\text{RIS}},m)^H \\
&+ \textbf{H}_{\text{rad}}^{\text{Tx-RIS}}(m,d),
   \end{split}
\end{equation} 
where $l_{\text{main}}^{\text{Tx-RIS}}(m,d)$ is the channel gain between the Tx and RIS for the frequency bin $m$, $\textbf{a}_{\text{RIS}}(\phi_{\text{Tx}},m) \in \mathbb{C}^{N_{\text{RIS}} \times 1}$ and $\textbf{a}_{\text{Tx}}(\theta_{\text{RIS}},m) \in \mathbb{C}^{N_{\text{Tx}} \times 1}$ denote the array steering vectors of the  RIS and the transmitter, directed to each other. Lastly, $\textbf{H}_{\text{rad}}^{\text{Tx-RIS}}(m,d)  \in   \mathbb{C}^{N_{\text{RIS}} \times N_{\text{Tx}}}$ denotes the re-radiation component of the THz channel, modelled as an interference to system, by evaluating eq. (\ref{eq:hint}) for each antenna pair.

Similarly, the channel from RIS to Rx, reflected over the target $k$ is denoted by $\textbf{H}_{\text{Rx,\textit{k}}} \in  \mathbb{C}^{N_{\text{Rx}} \times N_{\text{RIS}}}$ is given as
\begin{equation}
\begin{split}
\textbf{H}_{\text{Rx,\textit{k}}}[m] &=
l_{\text{main}}^{\text{RIS-\textit{k}-Rx}}(m,d)
\textbf{a}_{\text{Rx}}(\phi_k,m) 
\textbf{a}_{\text{RIS}}(\theta_k,m)^H\\
&+\textbf{H}_{\text{rad}}^{\text{RIS-\textit{k}-Rx}}(m,d),
\end{split}
\end{equation}
where,  $l_{\text{main}}^{\text{RIS-\textit{k}-Rx}}(m,d)$ denotes the combined channel gain for the given path. The term $\textbf{H}_{\text{rad}}^{\text{RIS-\textit{k}-Rx}}(m,d) \in   \mathbb{C}^{N_{\text{Rx}} \times N_{\text{RIS}}}$ denotes the re-radiation interference, similarly computed for each antenna pair from eq. (\ref{eq:hint}). Lastly, $\textbf{a}_{\text{Rx}}(\phi_k,m) $ and  $\textbf{a}_{\text{RIS}}(\theta_k,m)$ denote the steering vectors from the target $k$ to RIS and from target $k$ to Rx, respectively. The angles $\phi_k$ and $\theta_k$ are referred to as the angle-of-arrival (AoA) and angle-of-departure (AoD), respectively, which are variables must be determined for localization.




%% file: 2.2_diff.tex
The configuration given in Fig. \ref{fig:system} contains a blockage obstructing the line-of-sight. It is known from diffraction theory that the sharp edges of an obstacle may induce a secondary source of radiation. This phenomenon is called knife-edge diffraction and has been shown to be prominent in the terahertz band as well \cite{knife_edge}. In our case, this effect causes some signal power to leak behind the obstacle and interfere with the localization process.

\begin{figure}[ht]
    \centering
    \includegraphics[scale= 0.78]{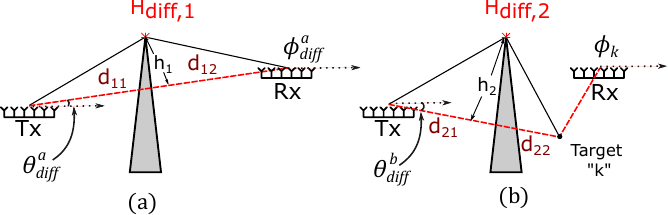}
    \caption{Two channels might occur due to diffraction. On the left, diffracted signal directly reaches the receiver. On the right, diffracted energy is reflected from the target before reaching the receiver. }
    \label{fig:diff_modes}
\end{figure}

The diffraction might affect the system performance in two main ways, as shown in Fig. \ref{fig:diff_modes}. First, the diffracted signals might directly get collected by the receiver. This is a harmful effect on the localization performance, as this signal contains no information regarding the targets in the shadowed area. Secondly, the signals diffracted may propagate towards the shadowed region and reflect off any objects present there. This path, on the contrary, is effectively contributes to the target echo in the receiver, therefore a helpful effect. Since the system performs localization by processing second-order reflections (first over the RIS and later over the target), both of these are as significant as the intended signal path, hence must be included in the system model.
The knife-edge diffraction phenomenon depends heavily on the problem geometry. The received power at a point in the shadowed region is calculated by assuming secondary Hugyen's sources above the blockage and calculating the total power reaching that point by taking Fresnel integrals \cite{rapapport}. As a result of this, an equivalent model for the LoS case is generated. To this end, some auxiliary geometrical variables must be defined. An imaginary line from the transmitter to the receiver is shown in Fig. \ref{fig:diff_modes}a. This line is divided perpendicularly into two parts as $d_1$ and $d_2$ by the line segment $\textit{h}$. The line segment $\textit{h}$ is referred to as the \textit{obstuction depth}. The geometric variables for the second diffraction scenario given in Fig. \ref{fig:diff_modes}b are also defined similarly. Based on these variables, the Fresnel parameter $\nu$ is defined as
\begin{equation} \label{eq:freshnel}
    \nu = \textit{h} \sqrt{\frac{2}{\lambda} \left( \frac{1}{d_1}+ \frac{1}{d_2}\right) },
\end{equation}
which is another auxiliary variable for calculating the ratio of electrical field strengths between the equivalent LoS channel and the actual path that the electromagnetic waves travel.
The knife-edge diffraction loss is given as
\begin{equation} \label{eq:diff}
  l(\nu)  = \left|       
  \frac{E_{\text{b}}}{E_{\text{u}}} \right| 
       = \sqrt{ \frac{(1 - C(\nu) - S(\nu))^2+(C(\nu)-S(\nu))^2}{4} },
\end{equation}
 where $|E_{\text{u}}|$ and $|E_{\text{b}}|$ represent the path gains for the free-space case and the diffracted case, respectively. $S(\cdot)$ and $C(\cdot)$ are Fresnel integrals \cite{rapapport}. The channel transfer function modeling the diffraction effects can be obtained by multiplying the point-to-point channel with the diffraction loss factor calculated in (\ref{eq:diff}). Specifically, for the case presented in Fig. \ref{fig:diff_modes}a the transfer function between the Tx and Rx is given as
 \begin{equation}
    \textbf{H}_{\text{Rx,Tx}}[m] = 
l_{\text{main}}^{\text{Tx-Rx}}(m,d)
    \textbf{a}_{\text{Rx}}(\phi_{\text{diff}}^a,m) 
    \textbf{a}_{\text{Tx}}(\theta_{\text{diff}}^a,m)^H
    +\textbf{H}_{\text{rad}}^{\text{Tx-Rx}},
 \end{equation}
 where, $\theta_{\text{diff}}^a$ and $\phi_{\text{diff}}^a$ denote the departure and arrival angles between the Tx-Rx channels, as depicted in Fig. \ref{fig:diff_modes}a. Further, the diffraction effect corresponding to this antenna and blockage combination becomes
 \begin{equation}
\textbf{H}_{\text{diff,1}} [m] = 
l(\nu_1^m)
\textbf{H}_{\text{Rx},\text{Tx}}[m],
\end{equation}
where the $\nu_1^m$ is obtained by plugging the geometric distances given in the Fig. \ref{fig:diff_modes}a to eq. (\ref{eq:freshnel}), for the frequency bin $m$. The transfer function of the second path shown in the Fig. \ref{fig:diff_modes}b is similarly defined as
 \begin{equation}
 \begin{split}
    \textbf{H}_{\text{Rx,k,Tx}}[m] &=
l_{\text{main}}^{\text{Tx-\textit{k}-Rx}}(m,d)
\sigma_k(m)
\textbf{a}_{\text{Rx}}(\phi_k,m)
\textbf{a}_{\text{Tx}}(\theta_{\text{diff}}^b,m)^H\\
&+\textbf{H}_{\text{rad}}^{\text{Tx-\textit{k}-Rx}},
\end{split}
 \end{equation}
 where $l_{\text{main}}^{\text{Tx-\textit{k}-Rx}}(m,d) = l_{\text{main}}^{\text{Tx-k}}(m,d)l_{\text{main}}^{\text{k-Rx}}(m,d)$ is the combined channel gain, obtained by evaluating eq. (\ref{eq:channel_gain}) with appropriate frequency and distance. And $\textbf{H}_{\text{rad}}^{\text{Tx-k-Rx}}$ is the terahertz re-radiation of the path, formulated by eq. (\ref{eq:hint}). Consequently, the knife-edge diffraction effect for this path can be written as
\begin{equation}
\textbf{H}_{\text{diff,2},\textit{k}}[m] = 
l(\nu_2^m)  
\textbf{H}_{\text{Tx}, k,\text{Rx}}[m] 
e^{j2\pi \beta_d},
\end{equation}
where $\nu_2^m$ is obtained from eq. (\ref{eq:freshnel}) by using the geometrical information in Fig.  \ref{fig:diff_modes}b for each frequency bin $m$. A random phase factor $e^{j2\pi \beta_d}$ is included in this formulation the model the any small displacements in the target position, where $\beta_d$ is a random variable uniformly distributed in $[0,1)$.

%% file: 3_algorithm.tex
The system achieves localization of the target by collaborative operation of the RIS and the radar. First, the transmitter directs the sensing signal to the known position of RIS. The RIS reflects the signals into the shadowed area, scanning a different angle sector at each transmission. The targets in the investigated angle sector get illuminated by the RIS, and the signals reflected from them arrive at the receiver. The AoA is estimated by direction-of-arrival (DoA) estimation methods at the receiver, and AoD is determined by the angle sector that RIS is focusing on. The AoA and AoD estimates are denoted by $\hat{\phi}_k$ and $\hat{\theta}_k$, respectively. Finally, the position of the target is determined by intersecting two lines passing through $\textbf{p}_{\text{RIS}}$ and $\textbf{p}_{\text{RX}}$ with angles $\hat{\theta}_k$ and $\hat{\phi}_k$ respectively.

\subsection{Hierarchical Codebook Design}
\begin{figure}[ht]
     \centering
     \includegraphics[scale=0.35]{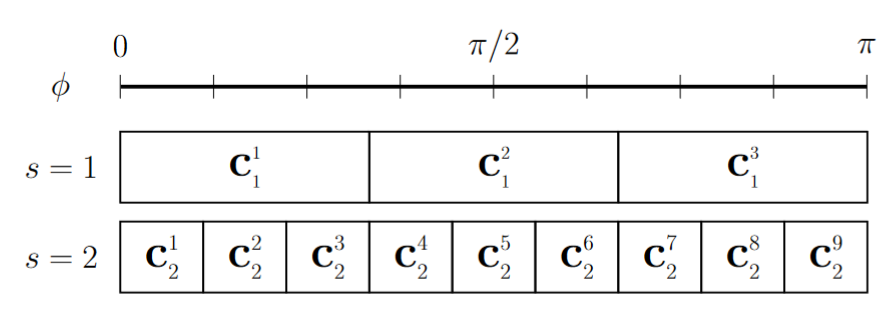}
     \caption{An example HCB with two stages and three codewords per stage. $(L= 3, S =2)$ }
     \label{fig:HCB}
 \end{figure}

 The performance of the localization heavily depends on the resolution of angle bins used at the RIS. Using sharp beams and scanning fine angle sectors at each transmission would enhance the performance since it would increase beamforming gain and decrease the ambiguity between the angle sector $\hat{\theta}_k$ and the true AoD of the target. However, this kind of exhaustive search would greatly increase the scanning time since it would require many transmissions to completely scan the scene. Even more, considering the targets are sparsely distributed along the scene, it is reasonable to assume that most of the transmissions would be wasted as no targets would get illuminated. A hierarchical codebook (HCB) search is proposed to solve this problem \cite{hcb09}. A hierarchical codebook is a set of codewords, i.e., beamformers, that is arranged hierarchically to scan the scene in an iterative fashion. The codebooks are designed to span the entire angular domain $[0,\pi]$ with equal-width beamformers. At each stage, the number of beamformers spanning the scene increases exponentially, making the resolution of the angle bins finer. An illustration of a hierarchical codebook is given in Fig. \ref{fig:HCB}, where $s$ denotes the stage index, and $L$ denotes the number of codewords per stage. Each stage contains $L^\text{s}$ codewords. In this example, $L$ is given as 3, and the total number of stages $S$ is given as 2, which means the whole angular sector
 $[0,\pi]$ is covered by three codewords in the first stage and nine codewords in the second stage. Each codeword covers the angle sector designated in the figure. All codewords must be designed to provide constant maximum directive gain in the angle sector they have assigned and minimum gain everywhere else. Moreover, since RIS is a passive device, these codewords must be phase-only beamformers. To this end, a set of analog beamformers with constant modulus are needed. In addition, RIS works as a phase shifter and cannot perform multiband processing, designing codewords for each frequency bin is impossible. Therefore, the codewords must be designed by taking a certain band as a reference. 

Assuming a uniform grid along the angular spectrum given by $\theta = \pi r/R$ for $r = 0, 1, \dots R$, the reflection beampattern of the RIS for our problem can be formulated by
\begin{equation}
    \text{BP}_{\text{RIS}}^l (\theta,m) =  
    \textbf{A}_{\text{RIS}}^\text{H} (\theta,m) 
    \bm{\Psi}
    \textbf{H}_{\text{Tx}}[m]\textbf{x}[m]
\end{equation}
where $\textbf{A}_{\text{RIS}}^\text{H} \in \mathbb{C}^{N_{\text{RIS}} \times R}$ is the array steering matrix, consisting of stacking the RIS array steering vectors for every angle in the grid. Ignoring the interference components in the THz function, which is impossible to design, the normalized beampattern can be approximately equivalent to
\begin{equation}
\begin{split}
    \text{BP}_{\text{RIS}} (\theta,m) \approx  
    \textbf{A}_{\text{RIS}}^\text{H} (\theta,m) 
    \bm{\Psi}
\textbf{a}_{\text{RIS}}(\phi_{\text{Tx}},m)
\end{split}
\end{equation}
The RIS HCB design problem is finding the codewords that yield certain beampattern at the central frequency, which can be achieved by parameterizing the beampattern as a function of $\bm{\Psi}$ evaluated at central frequency bin. Defining an auxiliary variable $ \textbf{c}_s^r = \bm{\Psi}_s^r \textbf{a}_{\text{RIS}}(\phi_{\text{Tx}})$, the codebook design problem can be formulated as
 \begin{equation} \label{eq:opt}
 \begin{aligned}
      \underset{\textbf{ c}_s^{r} }{\min} \quad &\| 
      \textbf{A}_{\text{RIS}}^\text{H}(\theta)
      \textbf{ c}_s^{r} - \textbf{d}_s^{r} \|_2,\\
          \text{s.t} \quad &|\textbf{c}_s^{r}(i)|^2 =  \frac{1}{\sqrt{N_{\text{RIS}}}}, i=1,\dots,N_\text{RIS}.
 \end{aligned}
 \end{equation}
 where $\textbf{d}_s^{r} \in \mathbb{C}^{R \times 1}$ is the ideal beam pattern of the RIS for the angular sector $r$. The structure of the $\textbf{d}_s^r$ for the first angle sector is $\textbf{d}^{1}_s =  [\textbf{1}_s^T \textbf{0}^T ]^T$  with $R/L^s$ ones and $R - R/L^s $ zeros. Note that the ideal beam pattern for each subsequent angle sector is $R/L^s$ entries circularly-shifted version of the previous vector. For this optimization problem, no closed-form solution exists due to the non-convex constraint in (\ref{eq:opt}). Hence, the gradient projection algorithm \cite{modulus} is adopted, i.e., alternately applying the gradient descent step and the projection of the solution onto a solution with constant modulus.
 
\subsection{Direction-of-Arrival Estimation at the Receiver} \label{chp:doa}

We can employ any direction-of-arrival estimation method to obtain the spatial spectrum at the receiver. Since the signal model is assumed to consist of multiple coherent frequency bands covering a wide spectrum as explained in Section \ref{chp:system}, multiband DoA methods must be utilized. As the signal subspace differs for different frequency bands, subspace methods necessitate a pre-processing method to create a consistent averaging of the signal subspace \cite{TUNCER}. To this end, we construct focusing matrices $\textbf{F}$ to project the subspace of each band to the central band such that
 \begin{equation} \label{eq:wbdoa}
 \begin{aligned}
\underset{\textbf{F}(f_m) }
{\min}
\quad &\| 
\textbf{A}_{\text{Rx}}(f_c,\theta)
-\textbf{F}(f_m)
\textbf{A}_{\text{Rx}}(f_m,\theta)
\|_F,\\
\text{s.t}
\quad
&\textbf{F}(f_m)^H
\textbf{F}(f_m)
=  \textbf{I}, 
\quad
i=1,\dots,M.
 \end{aligned}
 \end{equation}
where $\textbf{A}_{\text{Rx}}(f_c,\theta)$ is the array steering matrix for the Rx array, evaluated at the central frequency. The equation (\ref{eq:wbdoa}) has a closed-form solution, which is expressed as
$\textbf{F}(f_m) = 
\textbf{V}(f_m)
\textbf{U}(f_m)^H$, 
where the columns of  $\textbf{V}(f_m)$ and $\textbf{U}(f_m)$ are left and right singular vectors of  the product $\textbf{A}(f_c,\theta) \textbf{A}(f_m,\theta)^H$, respectively. As can be seen from the formulation, the solution of the focusing matrix requires a preliminary estimate of the DOA parameters, which can be obtained by using narrowband DoA estimation on the central frequency band \cite{wbdoa}. After obtaining the focusing matrices, the coherently averaged covariance matrix can be obtained by
\begin{equation}
    \hat{\textbf{R}}_{yy} =
    \sum_{m=1}^M
    \textbf{F}(f_m)
    \textbf{R}_{yy}(f_m)
    \textbf{F}(f_m)^H.
\end{equation}
After that, any kind of covariance matrix-based methods such as Capon beamformer, MUSIC, or ESPIRIT \cite{ASP} can be used to obtain DoA estimates. The subspace-based methods require robust covariance matrices to achieve their full potential. Therefore constructing the covariance matrix from multiple snapshots (say $B$ many) by $\textbf{R}_{yy} = \sum_{b=1}^B \textbf{y}_b\textbf{y}_b^H$ is recommended. Note that each stage must be repeated $B$ times here, which could slow the localization process but offer superior performance compared to the conventional beamformer, which relies on a simpler method of producing the spatial spectrum by evaluating the receive beampattern for all angles in the grid.
\cite{2decades}.

\subsection{Proposed Algorithm} \label{chp:prop_alg}
The transmitter sends probing signals toward the RIS. In each transmission, RIS first reflects the incoming signal to an angle sector set by the HCB to obtain the received signal $\Bar{\textbf{y}}$. In the subsequent transmission, the RIS is configured to reflect the signal back towards the transmitter, capturing the diffraction effects and producing the signal $y_d$ without target echoes. Then, the diffraction effects are subtracted to get the processed signal by $\textbf{y} =\Bar{\textbf{y}} - \textbf{y}_d $. The processed signal  ${\textbf{y}}$ is then put into a DoA algorithm, and the normalized spatial spectrum is computed. The spatial spectrum is compared to a threshold $\gamma$ to determine the possible angle bins that might contain targets. If the spectrum at the angle bin $\hat{\phi}_m$ is greater than $\gamma$, the codebook corresponding to that transmission is noted and investigated with refined codewords at the next stage of the HCB sweep. Any angle sector that does not give a peak exceeding the threshold is omitted during the next stage of the sweep, accelerating the localization process dramatically.

The threshold parameter $\gamma$ affects the dynamic range of the objects that the radar system can detect. The target detection is performed on the spectrum normalized to its maximum value. Therefore, the echoes of the closest target to the farthest one are mapped to range $(0,1]$. Choosing $\gamma$ high may cause distant targets to get discarded, whereas choosing it too low may cause false targets to appear in the system. The pseudo-code for the proposed algorithm is given in Algorithm 1.
\begin{algorithm}
\caption{Adaptive Multi-Target Localization}\label{alg:hcb}
\begin{algorithmic}
\State Set a threshold $0 < \gamma < 1$
\State Design the HCBs  $\textbf{C}_{\text{RIS}}^{(1)},\textbf{C}_{\text{RIS}}^{(2)}, \dots,\textbf{C}_{\text{RIS}}^{(S)}$
\State{Set the RIS profile: $\bm{\Psi}_{\text{Tx}}$}
\State{Collect calibration signal: $\textbf{y}_\text{d}$}
\State Set angle sector indexes to be scanned:\\ 
\quad {$\textbf{I}^{(0)} =[1,2,\dots,L]$}
\For{\textit{s} = 1,2,\dots, S}
    \For{\textit{l} = 1,2,\dots, $L^s$}
      \State{Set the RIS profile: 
      $\bm{\Psi}_l^s \leftarrow  
      \textbf{C}[\textbf{I}^{(s)}]$
      }
      \State{Transmit the probe signal $\textbf{x}$ at the Tx.}
      \State{Collect received signal: $\textbf{y}$ from the eq. (\ref{eq:y})}

    \EndFor
    \State{Perform background subtraction $\hat{\textbf{y}} =\textbf{y} - \textbf{y}_\text{d}$.}
    \State{Perform DoA and get pseudo-spectrum: 
    $\textbf{P}_{\hat{\textbf{y}}\hat{\textbf{y}}}$}
    \State{Choose the peaks $m  \leftarrow 
    \textbf{P}_{ \hat{\textbf{y}}\hat{\textbf{y}} }(m) \geq \gamma$}
    \State{Choose RIS phase shifts:}  
    \State{ \qquad $ 
          \textbf{I}^{(s)} \leftarrow  \underset{\Hat{l}} \argmax|  
        \textbf{P}_{
        \hat{\textbf{y}}\hat{\textbf{y}}} (\Hat{l})|
    $}
    \State{Calculate the HCB indexes for the next stage: }
     \For {p =  $\textbf{I}^{(s)}$}
        \State{$ \textbf{I}^{(s)} \leftarrow (p-1)L + 1:pL $}
     \EndFor
     \State{Send next HCB indexes $\textbf{I}^{(s)}$ to the RIS }
\EndFor
\State{Estimate the positions by intersecting AoA and AoD.}
\end{algorithmic}
\end{algorithm}

%% file: 3.1_refienment.tex
The performance of the algorithm described in the previous section is limited by the number of angle sectors in the codebook and DoA bins. This finite resolution produces ambiguous zones in the given algorithm and limits the performance of the system even in the high SNR regime. The size of the ambiguous zone and the corresponding error floor can be calculated straightforwardly in Appendix B. However, to obtain the maximum performance yielding the minimum error bound of the system,  a further refinement process is needed. 

The refinement process takes target angle estimates $\hat{\theta}_k$ and $\hat{\phi}_k$ as input, and produces RIS profiles directed near the vicinity of the target. For a  RIS configuration $\Psi^g$, the diffraction-reduced received signal reflected from a single target $k$ can be written as
\begin{equation}
\begin{split}
    \textbf{y}(m,g)  & = 
        \rho_k(m)
        \textbf{a}_{\text{Rx}}({\phi}_k,m) 
        \textbf{a}_{\text{RIS}}(\theta_k,m)^H
        \bm{\Psi}^g\\
        &\textbf{a}_{\text{RIS}}(\phi_{\text{Tx}},m)
        \textbf{a}_{\text{Tx}}(\theta_{\text{RIS}},m)^H
        \textbf{x}
        +
        \textbf{h}_{\text{rad}}^{\text{sys}}
        \text{x}\\
        &
        +\rho_{d_k}(m)
        \textbf{a}_{\text{Rx}}(\phi_{\text{k}},m)
        \textbf{a}_{\text{Tx}}(\theta_{\text{diff}}^b,m)^H
        e^{j2\pi \beta_d}
        \textbf{x}
        +\textbf{w}(m)
    \end{split}
\end{equation}
where $\rho_k(m) = \sigma_k(m) l_{\text{tot}}(m)$ denotes the total channel coefficient of the main signal path. Similarly, $\rho_{d_k}(m) = \sigma_k(m) l(\nu_2)  l_{\text{main}}^{\text{Tx-k-Rx}}(m)$ denotes the channel coefficient of the remaining diffraction effects. The term   $\textbf{h}_{\text{rad}}^{\text{sys}}$  models the combined re-radiation effects from all THz channels. 

The diffraction and the re-radiation terms contain random phase random parameters ($\beta_r$ and $\beta_d$) in which the distributions are known and independent of the position parameters $\theta_k$ and $\phi_k$. Consequently, these channels will not take part in the maximum likelihood (ML) estimate. In addition, the transmit signal is assumed to be known, as described in eq. (\ref{eq:tx_signal}). Since without loss of generality, the noise variance is assumed to be known, the remaining estimation parameters in the ML problem are the channel coefficient $\rho_k$ and the position parameter $\textbf{p}_k$. Therefore, the ML estimation problem can be formulated as
\begin{equation} \label{eq:argmle}
    \hat{\textbf{p}}_k^{\text{ML}} = 
    \arg \min_{\textbf{p}_k} 
    \left[ 
    \min_{\rho_k} 
    \textit{L} (\textbf{p}_k,\rho_k) 
    \right],
\end{equation}
where
\begin{equation} \label{eq:mle_3sum}
     \textit{L} (\textbf{p}_k,\rho_k)  = 
     \sum_{m=1}^M
     \sum_{k=1}^K 
     \sum_{g=1}^G
    \left \Vert
     \textbf{y}^g (m)
     -\rho_k(m)
     \textbf{A}^g(\textbf{p}_k,m)
     \textbf{x}(m)
    \right \Vert^2,
\end{equation}
where
\begin{equation}
\begin{split}
\textbf{A}^g(\textbf{p}_k(\theta_k,\phi_k),m) =&
        \textbf{a}_{\text{Rx}}({\phi}_k,m) 
        \textbf{a}_{\text{RIS}}(\theta_k,m)^H
        \bm{\Psi}^g\\
        &\textbf{a}_{\text{RIS}}(\phi_{\text{Tx}},m)
        \textbf{a}_{\text{Tx}}(\theta_{\text{RIS}},m)^H
    \end{split}
\end{equation}
is the term representing the main signal path. Now, for a given frequency bin $m$, the minimization problem inside the parenthesis of  eq. (\ref{eq:argmle}) is in least-squares form, therefore $\rho_k$ can be replaced by its least-squares estimate, namely
$\hat{\rho}_k^{\text{ML}} =
1/\sqrt{P_t}
\sum_{g=1}^G 
(\textbf{A}^g(\textbf{p}_k) 
\textbf{a}_{\text{Tx}}(\Tilde{\theta}_{\text{RIS}})
)^+ 
\textbf{y}^{g}$. 
In this case, the likelihood function becomes solely dependent on the position parameters. Since each frequency band is assumed to have no cross-interference between them, assuming other parameters are given, the signal becomes a Gaussian distributed over the frequency bin $m$. In that case, this parameter can be replaced by its ML estimate, the sample mean. Therefore, the estimation problem can be manifested as
\begin{equation}
    \hat{\textbf{p}}_k^{\text{ML}} = 
    \frac{1}{M}\sum_{m=1}^M
    \arg \max_{\textbf{p}_k } 
    \textit{L} (\textbf{p}_k,m),
\end{equation}
where    
\begin{equation} \label{eq:mle}
\begin{split}
   \textit{L} (\textbf{p}_k)  =  
   \sum_{k=1}^K 
   \sum_{g=1}^G
     ||\textbf{y}(m,g) 
    - & \hat{\rho}_k^{\text{ML}}
     \textbf{y}(m,g)\\
     &
     \textbf{A}^g(\textbf{p}_k,m) 
     \textbf{a}_{\text{Tx}}(\Tilde{\theta}_{\text{RIS}})
     ||^2.
\end{split}
\end{equation}
Note that on the one hand, due to the beam squint, a certain divergence from the RIS beamforming angle is expected at each frequency bin. On the other hand, the prediction model that we try to fit in eq. (\ref{eq:mle}) incorporates these effects naturally, at each frequency bin. The loss function given in eq. (\ref{eq:mle}) is highly non-convex and contains multiple local minima. Therefore, finding the correct parameters $\textbf{p}_k$ requires a detailed search in space. This search process can be drastically shortened by obtaining a good initial estimate and performing the search operation in the vicinity of this initial point. We can obtain good estimates by scanning the scene with HCBs as described in the previous section and performing an iterative search on the likelihood function around the target estimates, which constitutes Algorithm 2, described in the pseudo-code below.

\begin{algorithm}
\caption{Iterative Maximum Likelihood Estimation}\label{alg:mle}
\begin{algorithmic}
\State{Get the initial estimates $\textbf{p}_k$ from Algorithm 1.}
\State{Set an initial zone $\textbf{z}(x,y)$ for grid search.}
\State{Calculate the RIS profiles $\bm{\Psi}^g$ covering the zone}
\State{Send Tx signals, apply RIS profiles, receive $\textbf{y}_g$}
\For {i = 1,\dots N}
\State{Perform a grid search on $\textit{L}(\textbf{p})$, find minima $\textbf{p}_k^i$}
\State{Create a smaller zone centered around $\textbf{p}_k^i$ }
\EndFor
\end{algorithmic}
\end{algorithm}

The algorithm commences by defining a zone of investigation centered around the initial estimate obtained by the previous algorithm. After that, RIS phase profiles covering the whole region with 3dB separated beams are designed and transmitted sequentially. After transmitting all of the phase profiles, a 2D grid search is performed on the zone, evaluating the likelihood function at each point. Finally, a smaller, refined zone is constructed centered around the point, giving the minimum of the likelihood function. This process is performed multiple times, shrinking the estimation zone each time. In practice, it is observed that the algorithm is able to obtain position error bound (PEB) in less than 20 iterations by shrinking the zone by \%25 at each iteration. In the next section, we will derive the position error bound for the given system.


%% file: 4_crb.tex
The maximum achievable accuracy of the system is called position error bound and can be obtained by Cramer-Rao lower bound (CRLB) analysis. CRLB is a lower bound on the variance of an unbiased estimator, which is the best realizable estimator in terms of mean-square error \cite{SKay}. Therefore the mean square error of the position estimates obtained by any algorithm cannot be lower than this bound. In this problem, the unknown parameters for any target are position $(\theta_k, \phi_k)$ and the corresponding reflection coefficient $\rho_k$. Assume all unknown parameters of the all targets are stacked into a 3K-dimensional vector such as
\begin{equation} \label{eq:parameters}
      \bm{\eta}_{}  = \begin{bmatrix}\theta_1&\dots&\theta_K&\phi_1&\dots&\phi_K&\rho_1\dots\rho_K \end{bmatrix}^T,
 \end{equation}
In this case, the mean-square error between the true values $\bm{\eta}$ and their estimates $\bm{\hat{\eta}}$ are bounded by
\begin{equation}
    \mathbb{E} [(\bm{\hat{\eta}} - \bm{\eta} ) (\bm{\hat{\eta}} - \bm{\eta} )^H] \geq \textbf{J}^{-1}_{\bm{\eta}},
\end{equation}
where $\textbf{J}_{\bm{\eta}}$ denotes the Fisher information matrix (FIM). For our case, the Fisher Information matrix conveys the information that the receiver $\textbf{y}$ carries about the parameter vector $\bm{\eta}$. It is obtained by taking partial derivatives of the log-likelihood function with respect to each parameter. The $(p,q)$th entry of this matrix is given as
\begin{equation}
    [\textbf{J}_{\bm{\eta}}]_{p,q} = 
    \mathbb{E} 
    \left[
    -\frac{\partial^2 \ln{p}(\textbf{y}|\bm{\eta})}{\partial \eta_p \partial \eta_q} 
    \right],
\end{equation}
where $\ln{p}(y|\bm{\eta})$ denotes the log-likelihood function of $\textbf{y}$ given $\bm{\eta}$. Further, $\eta_p$  and $\eta_q$ denote the corrsponding variables in the vector $\bm{\eta}$. The log-likelihood distribution can be modeled as a multivariate symmetric complex Gaussian distribution, and consequently, the elements of the FIM become 
\begin{equation} \label{eq:crb}
    [\textbf{J}_{\bm{\eta}}]_{p,q} = 
    \sum_{m=1}^M
    \sum_{\xi=1}^\Xi
    \frac{2}{\sigma^2} 
    \operatorname{Re} 
    \left \{ 
    \frac{\partial \bm{\mu}^H (\xi) }{\partial \eta_p} \frac{\partial \bm{\mu}(\xi) }{\partial \eta_q } 
    \right \},
\end{equation}
where $\xi$ is the index of the total $\Xi$ transmissions, $m$ is the frequency band index, and
$\bm{\mu}(\xi) = \textbf{y} - \textbf{w} $ is the noise-free component of the received signal \cite{liuCRB}.\\

Note that the position parameters, namely the coordinates of the targets, are a function of angular parameters estimated in $\bm{\eta}$. We can obtain the FIM for the positions by applying a transformation to the FIM of the channel parameters, which can be written as
\begin{equation}
    \textbf{J}_{pos} =  \textbf{T} \textbf{J}_{\bm{\eta}_{\text{ch}}} \textbf{T}^T,
\end{equation}
where $\textbf{J}_{pos}$ denotes the FIM of the position parameters, and the matrix $\textbf{T}$ denotes the Jacobian transformation matrix, which conveys the geometrical relations between the angles and positions. The individual elements of both $\textbf{J}_{pos}$ and  $\textbf{T}$ are given in Appendix A.

Finally, the position error bound is found by taking the trace of the position-related entries of FIM \cite{liuCRB}. The first $2K$ entries of the parameter vector $\bm{\eta}$ are position parameters. Therefore the lower bound can be calculated as
\begin{equation}
    \text{PEB} = \sqrt{\text{tr} \left( \left [\textbf{J}_{\text{pos}}^{-1} \right ]_{1:2K,1:2K} \right) } .
\end{equation}
In summary, the PEB is a function of target parameters and SNR, combined at each frequency bin.

%% file: 5_results.tex
In this section, we present the studies that we conducted for the proposed system. In our simulations, we have considered a scenario where Tx, RIS, and Rx are located at $\textbf{p}_{\text{Tx}}=(2, 1)^T$, $\textbf{p}_{\text{RIS}}=(3.25, 0.25 )^T$ and $\textbf{p}_{\text{RX}}=(6.75, 0.25)^T$, respectively. The blockage is assumed to be a planar object, and for the experiments in Section \ref{chp:res}, it is placed parallel to the y-axis at coordinates $(5,0)$ to $(5,1.875)$. Moreover, the number of elements of each array are $\textbf{N}_{\text{Tx}}=256$, $\textbf{N}_{\text{RIS}}=64$ and $\textbf{N}_{\text{RX}}=128$. The operating frequency of the system is assumed to be $f_c = 300$ GHz. The transmit signal occupies a total bandwidth of $5$GHz, consisting of 4 equally-spaced coherent sub-bands of $10$MHz. The noise power collected with this bandwidth is determined by the formula $\sigma^2 = N_0B_w$, where $\sigma^2$ is the total noise power, $B_w$ is the bandwidth of the system, and the constant $N_0 = 10^{-14.4}$W/Hz is the noise spectral density. The bandwidth of each sub-band is chosen to ensure RCS flatness within each sub-band. For the results, the SNR is defined as the ratio of the transmitted signal power to the noise power in the receiver. Defining the SNR in this way helps us understand the required power of the signal at the transmitter to achieve a reasonable localization performance. 

Two different types of HCBs have been designed with $L = \{ 5, 7 \}$ and $S = \{3, 3\}$. At the receiver, DoA estimation is performed on a uniform grid with $\pi/N_{\text{Rx,res}} = \pi/180$ resolution. The sample root mean square error (RMSE) was calculated using 100 estimations to approximate the RMSE performance.

\subsection{Blockage and Impact of Diffraction}
According to the model given in Section \ref{chp:system}, the path loss due to the diffraction is heavily dependent on the system geometry. Therefore, it is important to understand the impact of the diffraction at different setups. The knife-edge loss is a ratio normalized to the unblocked case. When the eq. (\ref{eq:diff}) is evaluated for different values of $h$, it is observed that the effect of the diffracted channel rapidly decreases as the transmitter is hidden deep behind the blockage.
\begin{figure}[ht]
    \centering
    \includegraphics[scale=0.5]{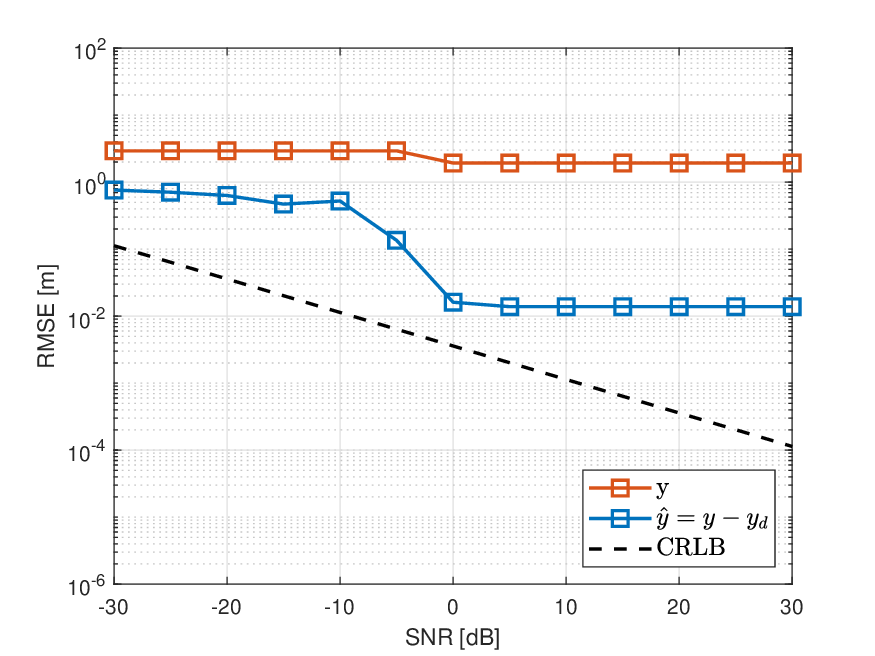}
    \caption{Impact of diffraction on positioning performance. The blue curve employs diffraction compensation. The red curve directly uses the received signal without diffraction compensation.
     }
    \label{fig:diffs}
\end{figure}
It can be argued that since the diffraction loss is quite high for any practical obstruction depths, both of the diffraction links must be ignored. However, the intended signal path drawn in Fig. \ref{fig:system} is actually a second-order reflection, and its signal model includes the multiplication of three distance-dependent loss terms, as given in the eq. (\ref{eq:y}). On the other hand, the diffraction link given in Fig. \ref{fig:diff_modes}a includes only a single virtual signal path, and the one given in Fig. \ref{fig:diff_modes}b includes a single reflection after the diffraction. Therefore, often, the diffraction links are as strong as the intended signal path. Consequently, the subtraction of the diffracted links becomes important. This is achieved by configuring the RIS to collect signals $\textbf{y}$ and $\textbf{y}_d$ at each stage, as explained in Section III-C. The impact of diffraction on localization performance can be seen in Fig. \ref{fig:diffs}, where the RMSE of localization for a system with HCB $(L,S)=[7,3]$  in a scene with a single target located at $\textbf{p} = (5.25, 3.75)^T$ is given. The experiment is done with both received signal $\textbf{y}$ and the diffraction-compensated signal $\Hat{\textbf{y}} =\textbf{y} -\textbf{y}_d$. Clearly, the uncompensated system is not able to obtain good performance even in high-SNR regimes.

\begin{figure*}[ht]
    \centering
  \subfloat[ Position estimates at -40dB. \label{fig:dif-40dB }]{%
        \includegraphics[scale=0.40]{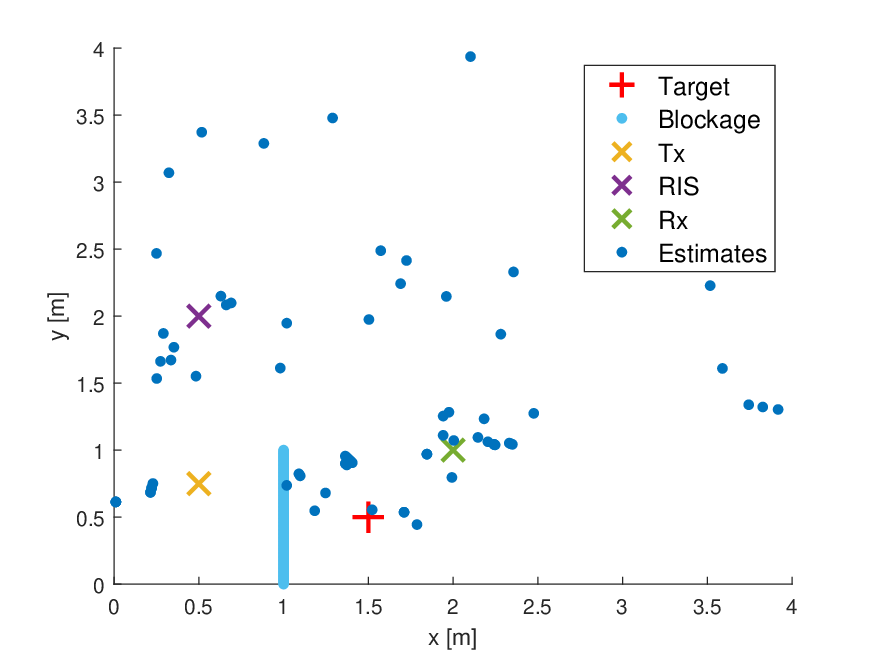}}
       \hfill
  \subfloat[ Position estimates at 0dB. \label{fig:dif 0dB}]{%
        \includegraphics[scale= 0.40]{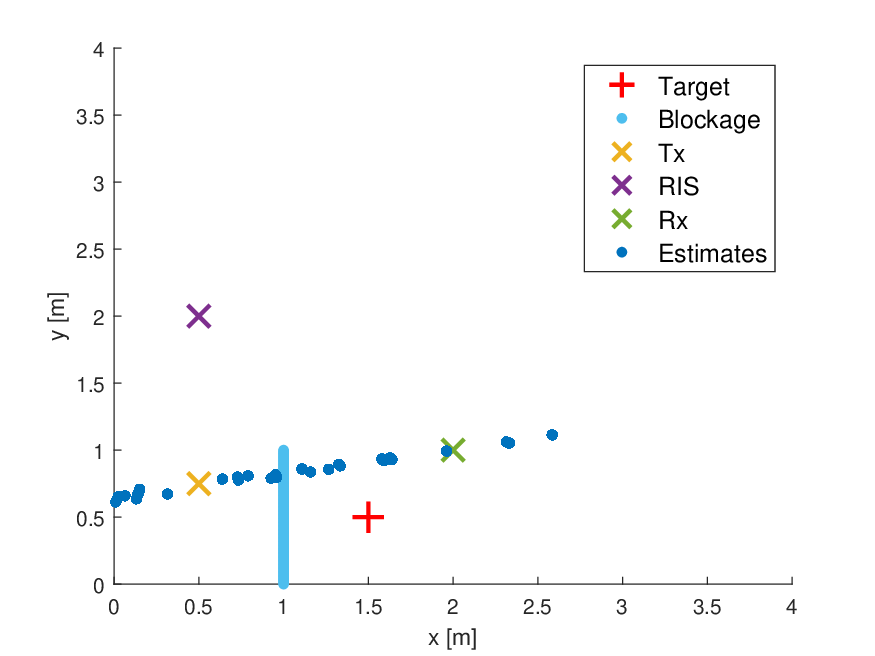}}
      \subfloat[ Position estimates at 40dB. \label{fig:dif40dB }]{%
       \includegraphics[scale= 0.39]{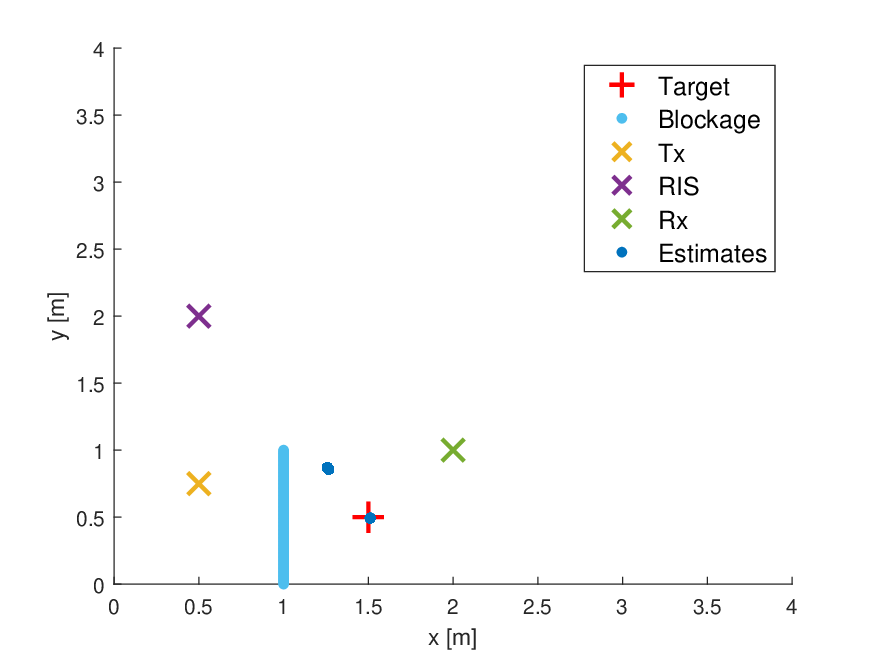}}
    \caption{The behavior of the system when diffraction effects are not subtracted. As the SNR increases, the diffracted signals dominate the AoA estimation at the receiver and cause incorrect estimates along the Tx-Rx line. }
  \label{fig:dif_points}
\end{figure*}

The reason for this behavior requires an explanation. The target estimates of the system for uncompensated signal are given in Fig. \ref{fig:dif_points}, where the setup is drawn on a 2D-coordinate system, and the positions of all elements of the system are shown. The obstruction is drawn as a cyan line, and the target estimates are plotted as blue dots. At low SNRs, the signal power of both intended and diffracted links is too low, and noise dominates the receiver. At this level, the localization is completely random, as shown in Fig. \ref{fig:dif-40dB }. As SNR increases, the signal power overcomes the noise. However, since the first diffracted link is the strongest path, this channel dominates the AoA estimation at the receiver, and the detections are concentrated along a line from the transmitter to the receiver, which can be shown in Fig. \ref{fig:dif 0dB}. Note that at this level, AoD estimation at the RIS is not completely random, as most of the detections are cluttered over the line passing from RIS and the target. The AoD estimation becomes precise with high SNRs, as can be seen in Fig. \ref{fig:dif40dB }, where almost all of the detections are placed at the same point, at the intersection of Tx-Rx and RIS-target lines. However, this estimation has huge bias due to the diffraction path and causes RMSE to stay roughly constant during all SNR regimes. In contrast, when the diffraction links are subtracted, no error occurs at AoA estimation, and the system performs accurately.\\
Note that the proposed diffraction cancellation method is not perfect, as it cannot cancel but might even amplify the second-diffraction path. However, in most practical cases, the first diffraction path is stronger since it is a first-order reflection. Consequently, this method is a simple and effective way to make the system work under these circumstances. Therefore, all of the results provided in the following sections are achieved after performing diffraction path cancellation.

\subsection{Hierarchical Codebooks vs Fully-Directive Scan: A Tradeoff} \label{chp:gamma}
The primary benefit of employing hierarchical codebooks over scanning the entire area with the most directive beam lies in the shortening of the search process. Terahertz arrays are capable of generating exceptionally narrow beams; for instance, the RIS simulated in this paper has a beamwidth of approximately $1.79^{\circ}$. Given that point targets are typically sparsely distributed across a few angle bins, if a non-hierarchical approach is employed, the entire search process would consist of a lengthy sequence of transmissions, resulting in no targets in most cases. Therefore, HCBs aim to trade directivity for search overhead. Consequently, the loss in directivity corresponds to a loss in the received signal strength, which might degrade the performance. Therefore, a study of detector threshold $\gamma$ and system performance must be conducted.

\begin{figure}[ht] \label{fig:gamma}
    \centering
  \subfloat[\label{fig:gamma_mse}
Localization performance comparison of fully-directive scan vs a 3-stage codebook.]{%
       \includegraphics[scale=0.48]{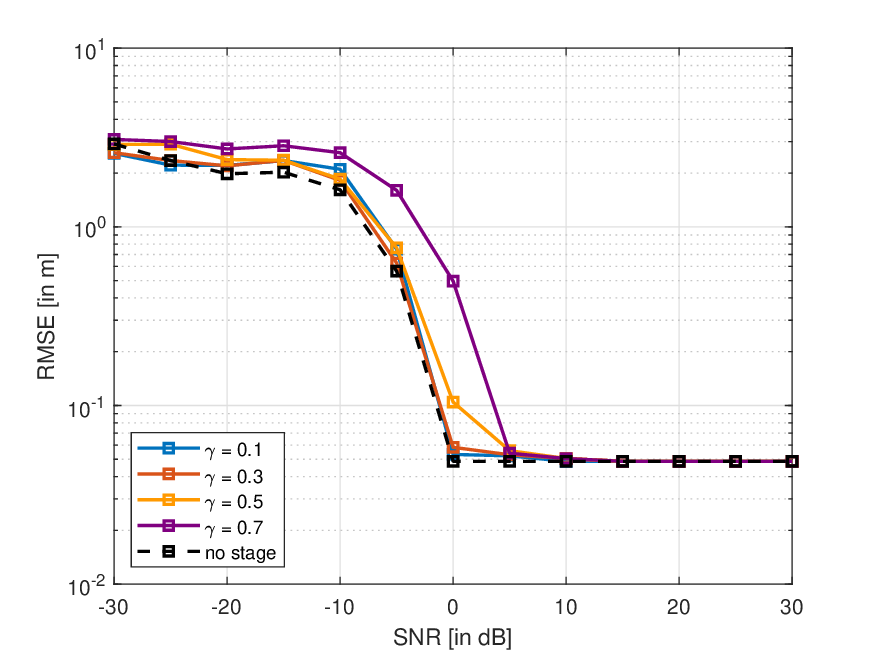}}
    \hfill
  \subfloat[\label{fig:gamma_det}
  Average number of transmissions required at each SNR to obtain the result in Fig.\ref{fig:gamma_mse}.]{%
        \includegraphics[scale= 0.48]{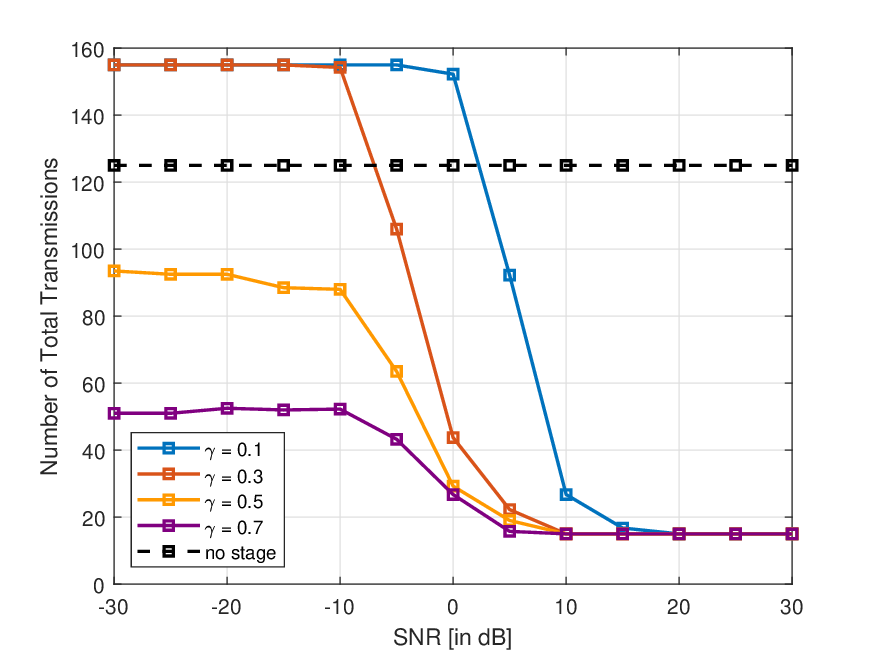}}   \\
    \caption{ The trade-off between HCB and full scan. The HCBs offer significant  reduction in the total transmissions in exchange for slight degradation in performance in low SNRs}
\end{figure}

In Fig. \ref{fig:gamma_mse}, the change of performance of the system using a $L^S = 5^3$ HCB in a scene with one point target at position $\textbf{p} = (5.25, 3.75)^T$ for different values of $\gamma$ is given. The performance of the single-stage search for the same resolution ($5^3=125$ angle sectors) is added as the baseline. It is observed that, an increasing threshold parameters causes a drop in localization performance. The reason behind this is that in low SNRs, some targets get undetected in the first levels of the codebook, where the provided beamforming gain is low. Consequently, the performance of the fully directive scan is slightly better, wherein in each transmission, the beam is focused on each angle sector sequentially. However, the advantage of the HCBs is clearly demonstrated in Fig. \ref{fig:gamma_det}, where, the  average number of transmissions for achieving the aforementioned localization performance is plotted. Full-directive search performs a transmission for each angle sector, totaling 125, regardless of the SNR. Meanwhile, the average number of transmissions of the HCBs depends on SNR and the threshold. It drastically drops with the SNR requiring only 15 transmissions for localizing a single target. The threshold parameter $\gamma$ affects the total transmissions in the low SNR regime, where noise dominates the system, causing many false alarms and requiring empty angle sectors to be inspected in the following levels. When $\gamma$ is chosen too low, the total transmissions could be even more than the full scan, which was the case for $\gamma = 0.1$ and $\gamma = 0.3$ in the Fig. \ref{fig:gamma_det}, where all $\prod_n^35^n = 155$ angle sectors ended up getting inspected. In conclusion, threshold parameter selection plays a role in achieving fast scanning of the shadowed area. Judging from the experiment conducted, $\gamma = 0.3$ provides a nice trade-off, where only a fraction of transmissions are required at the cost of a slight performance degradation at $0$dB.
\subsection{Single-Target Localization} \label{chp:single_target}
\begin{figure}[]
    \centering
  \subfloat[ Estimation mismatch of the target positions. \label{st_a}]{%
       \includegraphics[scale= 0.5]{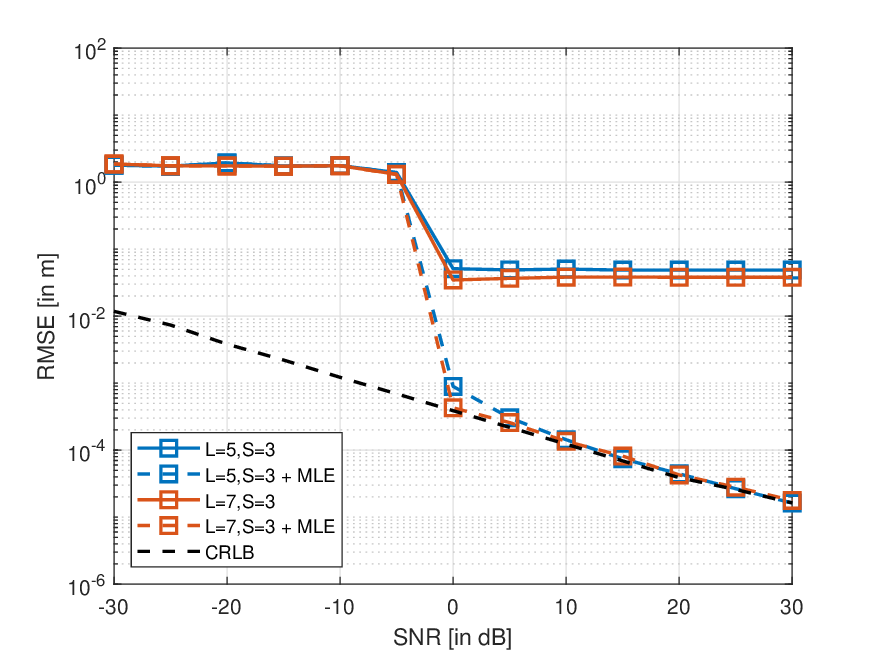}}
       \hfill
  \subfloat[ Average number of declarations ’target present’. \label{st_b}]{%
        \includegraphics[scale= 0.5]{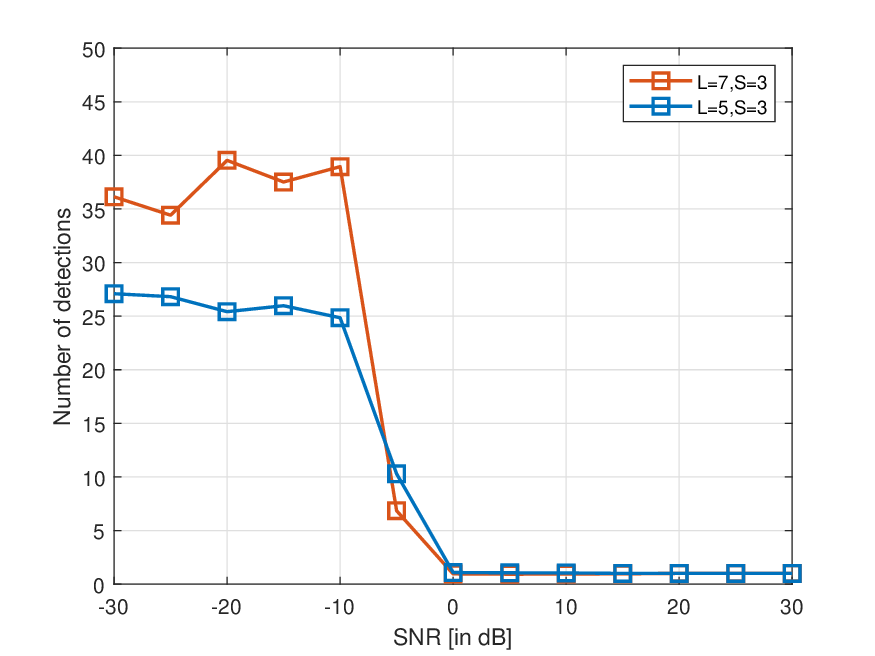}}
    \\
    \caption{(a) Single-target sensing performance of HCBs and additional refinement algorithm. (b) Average number of detections vs SNR.} 

  \label{fig:single_target} 
\end{figure}

For the single-target scenario, we consider a point target at position $\textbf{p} = (5.25, 3.75)^T$. The performance of the HCBs is evaluated at different levels of SNR while using the conventional beamformer at the receiver. The RMSEs of the estimated target positions for different SNR values are given in Fig. \ref{st_a}. At low SNRs, the spatial spectrum is dominated by noise, which causes false targets to appear and decrease the system performance. The system performance rapidly increases after $0$dB SNR and achieves centimeter-level accuracy. In general, it can be expected that the localization performance improves with higher codebook resolution, which is also observed in Fig. \ref{st_a}. The red curve with $L^S = 7^3$ codewords has a considerably lower error than blue $L^S = 5^3$ curves, particularly at high SNR regime. The performance of the codebooks saturates at the high SNRs due to limited angle sector resolution. For a known target position, this error floor can be analytically calculated, as shown in Appendix B. However, this method alone yields results significantly below the theoretical capabilities of a terahertz positioning system, as evidenced by the CRLB for the given problem. The additional ML refinement is able to achieve this impressive performance, obtaining sub-millimeter accuracy in a high SNR regime with a few extra transmissions. We observe that using denser codebooks helps obtain the CRLB performance sooner since better initial estimates are obtained with sharper beams.

The change of declared detections with SNR is given in Fig. \ref{st_b}. It is observed that at the low SNR regime, noise in the normalized spectrum causes many false targets to appear. However, improving SNR signal power rapidly exceeds the noise floor, and the system is able to detect only the target present in the scene. We observe that, in a low SNR regime, where noise dominates the spectrum, for a given threshold $\gamma$, the denser codebook $\{7,3\}$ tends to provide more false alarms due to having more angle sectors.

\begin{figure}[ht]
    \centering
  \subfloat[\label{fig:multi_target_a}
  Cumulative estimation mismatch of the target positions.  ]{%
       \includegraphics[scale=0.5]{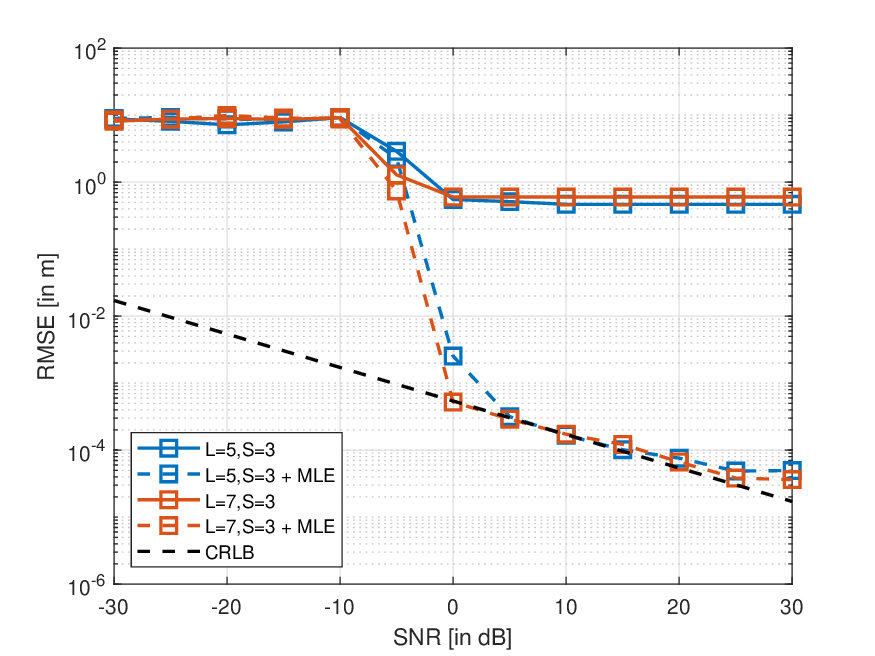}}
    \hfill
  \subfloat[\label{fig:multi_target_b}
  The average number of declarations ’target present’.]{%
        \includegraphics[scale= 0.5]{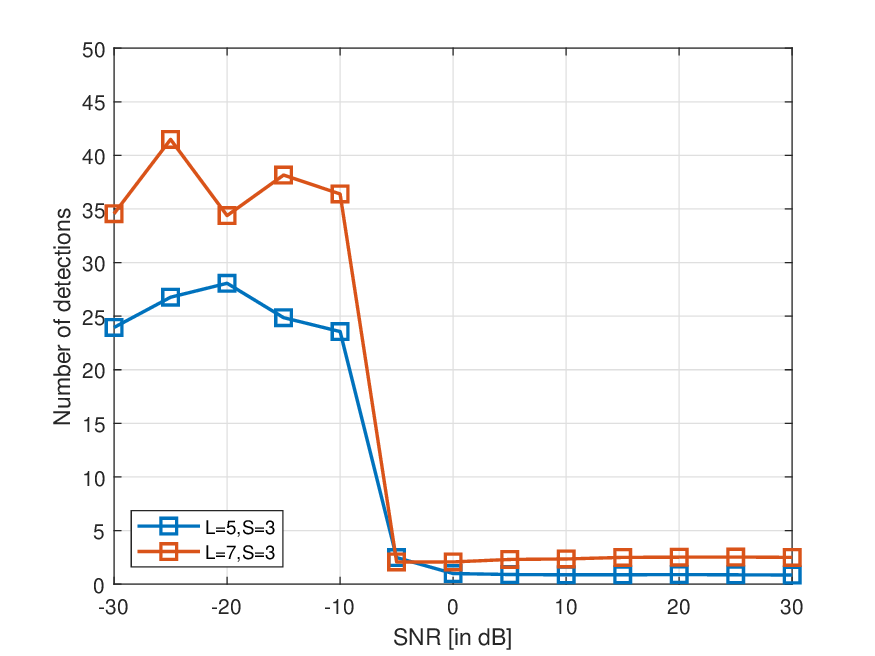}}   \\
    \caption{(a) Multi-target sensing performance of HCBs and following refinement algorithm. (b) Average number of detections vs SNR.}
\end{figure}
\subsection{Multi-Target Localization} \label{chp:multi_target}
For the multi-target localization experiment, we simulated a scene with the same system settings and three-point targets located at coordinates $\textbf{p}_1 = (4,4 )^T$,  $\textbf{p}_2 = (5,4)^T$, and  $\textbf{p}_3 = (6,4)^T$ respectively. The RMSEs of the estimated target positions are illustrated in Fig. \ref{fig:multi_target_a} for the three HCB designs across various SNR levels. In general, the performance is comparable to the single-target case, obtaining centimeter-level accuracy with the HCBs around $0$dB. Note that since this value is the total RMSE error for all of the targets, the positioning error for each target is even lower.
Similarly, the performance lower bound is obtained after the ML refinement process. The huge path losses of terahertz channels show their effect in this experiment. The losses in this band cause the echoes from the targets to rapidly diminish with the distance. This makes tuning the detection threshold parameter $\gamma$ critical, as it practically determines the range the system can detect, as discussed in Section \ref{chp:prop_alg}. The chosen $\gamma = 0.3$, gives the best trade-off between the scan time and the performance, as evaluated in Section \ref{chp:gamma}. Choosing a lower threshold parameter extends the dynamic range of targets that could be detected by the system but causes more false targets to appear, extending the scan time. In Fig. \ref{fig:multi_target_b}, the average number of detections is plotted for different SNR values. Note that for both single-target and multi-target cases, the average number of detections is not related to HCB configuration. The main factors effecting false-alarm rate is the threshold parameter $\gamma$ and the Rx resolution $N_{\text{Rx,res}}$. While $\gamma$ causes ghost targets to appear by lowering the detection level, $N_{\text{Rx,res}}$ effects increases the false alarms simply by increasing the number of angle bins to test. Conclusively, if empirical studies of false alarm rates are needed, appropriate normalizations must be applied to the statistics.

\subsection{Angular Diversity Provided by the Beam Squint}
\begin{figure} [ht]
    \centering
    \includegraphics[scale = 0.5]{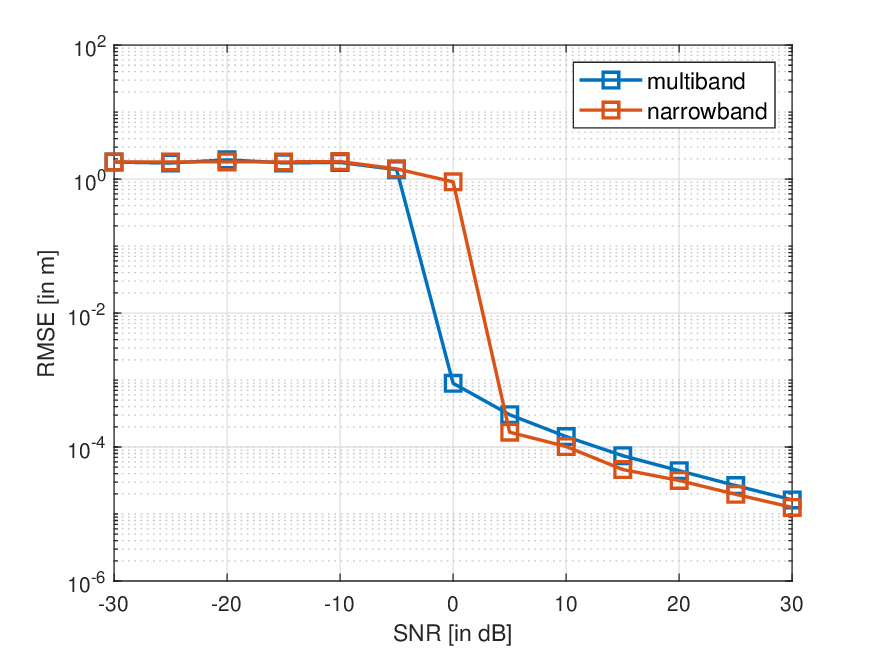}    \caption{Comparison of single band and multiband performance}
    \label{fig:widebandBS}
\end{figure}

The frequency difference between the sub-bands  causes a dispersion in the directed energy, also known as the beam squint. Although modern RF electronics can compensate for this effect using technologies such as true-time delays \cite{timedArrays}, RIS cannot do the same since they consist solely of passive phase-shifting elements. Consequently, beams reflected from the RIS will be slightly separated when they reach the target based on the sub-band. This separation introduces angular diversity by enabling slightly different illumination angles to the target, which are averaged over the sub-bands as described in the eq. (\ref{eq:mle_3sum}), thereby improving estimation performance. This improvement is illustrated in Fig. \ref{fig:widebandBS}, where RMSE performance for the scenario described in section\ref{chp:single_target} is simulated with a multiband configuration consisting four sub-bands of $10$MHz bandwidth equally distibuted across a $5$GHz band, versus a single sub-band of the same sub-band bandwidth, located at carrier frequency. Around the medium SNR regime (0-10dB), the multiband configuration demonstrates superior RMSE performance by providing more estimates to mitigate noise.

%% file: 6_conclusions.tex
In this work, we have proposed a sensing scheme for observing targets under NLOS conditions by a bistatic MIMO radar operating at the terahertz frequencies. We have demonstrated the effects of the terahertz signal propagation and diffraction on the sensing performance and derived a lower bound for the proposed system. We have shown that electromagnetic effects, which are so far ignored in RIS-aided radar setups, might radically hamper the performance of the system, and we proposed a method to attain the desired performance. In addition, to achieve fast scanning of the blocked area, we proposed a codebook-based search and subsequent refinement algorithm. Our results show that with a suitable HCB design and following the iterative refinement proposed, the ultimate sub-millimeter level sensing accuracy can be obtained.

Our future work will encompass a deeper investigation of beam squint and its usefulness as a method for area scanning, determining optimal number of sub-bands and their bandwidth as well as exploring other types of codebook schemes to achieve faster initial estimates.

%% file: Appendix_A.tex
Dropping the frequency index $m$ and the target index $k$ for clarity, the elements of the Fisher information matrix given in (\ref{eq:crb}) is defined as
\begin{equation}   
\begin{split}
    \frac{\partial [\mu]_l}{\partial \rho} =  
    &(\textbf{a}_{\text{RIS}}(\phi)
    \textbf{a}_{\text{Rx}}(\theta)^H \Psi
    \textbf{a}_{\text{RIS}}(\phi_{\text{Tx}})
    \textbf{a}_{\text{Tx}}(\phi_{\text{RIS}})^H +
    \\
    &l(\nu_2)
    l_{\text{main}}^{\text{Tx-k}}/(l_{\text{main}}^{\text{Tx-RIS-k}})
    \textbf{a}(\Tilde{\phi}_k) 
    \textbf{a}(\Tilde{\phi}_{\text{Tx}}) ^H
    e^{j2\pi \beta}
    )\textbf{x}
\end{split}
\end{equation}
\begin{equation}
\begin{split}
    \frac{\partial [\mu]_l}{\partial \phi} = 
    &   \frac{
        \partial \textbf{a}_{\text{RIS}}(\phi)}
        {\partial \phi}
    (
    \rho
    \textbf{a}_{\text{Rx}}(\theta)^H \Psi
    \textbf{a}_{\text{RIS}}(\phi_{\text{Tx}})
    \textbf{a}_{\text{Tx}}(\phi_{\text{RIS}})^H \\
    +
    &
    \rho_{d}
    \textbf{a}(\alpha_\text{Tx})^H e^{j2\pi \beta_d}
     \textbf{a}_{\text{Tx}}(\theta_{\text{RIS}})^H e^{j2\pi\beta_d} 
     )
     \textbf{x}
\end{split}
\end{equation}
\begin{equation}
    \frac{\partial [\mu]_l}{\partial \theta} = 
    \rho 
    \textbf{a}_{\text{Rx}}(\phi)
    \frac{
     \partial \textbf{a}_{\text{RIS}}(\theta)^H }
    {\partial \theta} \Psi
    \textbf{a}_{\text{RIS}}(\phi_{\text{Tx}})
    \textbf{a}_{\text{Tx}}(\phi_{\text{RIS}})^H \textbf{x},
\end{equation}
where
\begin{equation}
  \frac{
  \partial \textbf{a}_{\text{Rx}}(\phi)}
  {\partial \phi } = 
  \textbf{a}_\text{Rx}(\phi) \odot \left(-j 
  \frac{
  \partial \textbf{k}(\phi)}
  {\partial \phi} \right)
\end{equation}
\begin{equation}
  \frac{
  \partial \textbf{a}_{\text{RIS}}(\theta)}
  {\partial \theta } = 
  \textbf{a}_\text{RIS}(\theta) \odot \left(-j 
  \frac{
  \partial \textbf{k}(\theta)}
  {\partial \theta} \right),
\end{equation}
where the auxiliary variable $\textbf{k}$ corresponds to the phase shifts in a uniform linear array whose partial derivatives are defined as
\begin{equation}
  \frac{
  \partial \textbf{k}(\theta)}
  {\partial \theta } =  -\frac{2\pi d}{\lambda} \sin{\theta}
\end{equation}
\begin{equation}
  \frac{
  \partial \textbf{k}(\phi)}
  {\partial \phi } =  -\frac{2\pi d}{\lambda} \sin{\phi}.
\end{equation}
The partial derivatives contained in the Jacobian matrix are given as
\begin{equation}
    \textbf{T} =   \begin{bmatrix} 
\frac{\partial \phi}{\partial x}&
\frac{\partial \phi}{\partial y}&
\frac{\partial \phi}{\partial \rho_r}&
\frac{\partial \phi}{\partial \rho_i}\\
\\
\frac{\partial \theta}{\partial x}&
\frac{\partial \theta}{\partial y}&
\frac{\partial \theta}{\partial \rho_r}&
\frac{\partial \theta}{\partial \rho_i}\\
\\
\frac{\partial \rho_r}{\partial x}&
\frac{\partial \rho_r}{\partial y}&
\frac{\partial \rho_r}{\partial \rho_r}&
\frac{\partial \rho_r}{\partial \rho_i}\\
\\
\frac{\partial \rho_i}{\partial x}&
\frac{\partial \rho_i}{\partial y}&
\frac{\partial \rho_i}{\partial \rho_r}&
\frac{\partial \rho_i}{\partial \rho_i}\\ 
\end{bmatrix} = 
 \begin{bmatrix} 
 a & b & 0 & 0 \\
 c & d & 0 & 0 \\
 0 & 0 & 1 & 0\\
 0 & 0 & 0 & 1
\end{bmatrix}.
\end{equation}
Defining two auxiliary vectors $\textbf{u} = \textbf{p}_k - \textbf{p}_{\text{RIS}}$ and $\textbf{v} = \textbf{p}_k - \textbf{p}_{\text{Rx}}$, the elements of the Jacobian matrix is given as 
\begin{equation} 
\begin{aligned}
   &a = \frac{-1}{\sqrt{1-(\sfrac{[\textbf{v}]_1}{\|\textbf{v}\|})^2}}
       \qquad
    b = \frac{1}{\sqrt{1-(\sfrac{[\textbf{v}]_2}{\|\textbf{v}\|})^2}}\\ 
    &c =\frac{-1}{\sqrt{1-(\sfrac{[\textbf{u}]_1}{\|\textbf{u}\|})^2}} \qquad
    d = \frac{1}{\sqrt{1-(\sfrac{[\textbf{u}]_1}{\|\textbf{u}\|})^2}}\\ 
     \end{aligned}.
\end{equation}

%% file: Appendix_B.tex
Assume that positions of Rx and RIS are denoted $\textbf{p}_{\text{Rx}}, \textbf{p}_{\text{RIS}} \in \mathbb{R}^{2x1}$, respectively. Also, let $\textbf{u}$ and $\textbf{v}$ be the unit vectors pointing true AoA and AoD with headings  $\phi$ and $\theta$  respectively. If a target has been found by intersecting AoA and AoD, the following holds:
\begin{equation} \label{eq:IOL}
    \textbf{p}_{\text{Rx}} +c_1\textbf{u} = \textbf{p}_{\text{RIS}} +c_2\textbf{v},
\end{equation}
where, $c_1$ and $c_2$ are two constants. Defining $\textbf{q} = \textbf{p}_{\text{Rx}} - \textbf{p}_{\text{RIS}} $  and rearranging, the equation can be written as $\textbf{q} = \textbf{W} \textbf{c}$ which has the form
\begin{equation}
\begin{bmatrix}
    q_1\\
    q_2
\end{bmatrix}
= 
\begin{bmatrix}
    u_1 & -v_1\\
    u_2 & -v_2
\end{bmatrix}
\begin{bmatrix}
    c_1\\
    c_2
\end{bmatrix},
\end{equation}
where $u_1 = \cos{\phi}$,  $u_2 = \sin{\phi}$, $v_1 = \cos{\theta }$, and $v_2 = \sin{\theta}$, respectively. Since $\textbf{W}$ is full rank, we can invert the matrix of unit vectors to obtain constants
\begin{equation}
\begin{bmatrix}
    c_1\\
    c_2
\end{bmatrix} = 
\begin{bmatrix}
    u_1 & -v_1\\
    u_2 & -v_2
\end{bmatrix}^ {-1}
\begin{bmatrix}
    q_1\\
    q_2
\end{bmatrix}.
\end{equation}
 After determining the constants, the true location of the target can be found by evaluating either side of eq. (\ref{eq:IOL}), such as
 \begin{equation} \label{eq:ang2pos}
    \textbf{p}_{\text{target}} = c_1 \textbf{u} + \textbf{p}_{\text{Rx}}.
\end{equation} 
 
 Now, let $\hat{\theta} = \theta +\Delta_\theta$ and  $\hat{\phi} = \phi +\Delta_\phi$ denote the estimated AoD and AoA angles with $\Delta_\theta$ and $\Delta_\phi$ denoting the error terms. The effect of these error terms on the position estimate $\hat{\textbf{p}}_{\text{target}}$ would be
\begin{equation}
    \hat{\textbf{p}}_{\text{target}} = \hat{c}_1 \hat{\textbf{u}} + \textbf{p}_{\text{Rx}} = c_1 
    \begin{bmatrix}
        \cos(\phi + \Delta_\phi)\\
        \sin(\phi + \Delta_\phi)
    \end{bmatrix} +\textbf{p}_{\text{Rx}},
\end{equation} 
where
\begin{equation}
    \hat{c}_1 = \frac{(-q_1\sin{(\theta+\Delta_\theta)} + q_2\cos{(\theta + \Delta_\phi)})}{\sin{(\phi-\theta + \Delta_\theta-\Delta_\phi )}} .
\end{equation}

At high SNRs, the  main source of the error is angle beam resolution at RIS and Rx, which means the angular error terms are bounded by
\begin{equation} \label{eq:ang_err}
\begin{aligned}
    &\Delta_\theta \geq \theta  \bmod{  \left({\pi}/{2L^S} \right)}\\
    &\Delta_\phi \geq \phi  \bmod{ \left({\pi}/{2N_{\text{Rx,res}}} \right)}.
\end{aligned}
\end{equation}
Therefore, one can calculate the position error bound for the high SNR regime by assuming equality in eq. (\ref{eq:ang_err}) and calculating the RMSE between eq. (38) and eq. (39). 

For unknown target positions, a statistical model might be constructed by assuming the angular error terms are uniformly distributed at $\Delta_\theta \in [0,{\pi}/{2L^S}] $ and $\Delta_\phi \in [0, {\pi}/{2N_{\text{Rx,res}}}]$ and calculating the RMSE of target positions as before.